\documentclass{aa}  
\usepackage{graphicx}
\usepackage{setspace}
\usepackage{amssymb}
\usepackage{longtable,lscape}
\usepackage{rotating}
\usepackage{natbib}
\usepackage{color}
\usepackage{natbib}
\bibpunct{(}{)}{;}{a}{}{,}

\usepackage{graphicx}
\usepackage{txfonts}
\begin{document}

\newcommand{\FeII}{[\ion{Fe}{ii}]}
\newcommand{\TiII}{[\ion{Ti}{ii}]}
\newcommand{\SII}{[\ion{S}{ii}]}
\newcommand{\OI}{[\ion{O}{i}]}
\newcommand{\OIp}{\ion{O}{i}}
\newcommand{\PII}{[\ion{P}{ii}]}
\newcommand{\NI}{[\ion{N}{i}]}
\newcommand{\NII}{[\ion{N}{ii}]}
\newcommand{\NIp}{\ion{N}{i}}
\newcommand{\NiII}{[\ion{Ni}{ii}]}
\newcommand{\CaIIp}{\ion{Ca}{ii}}
\newcommand{\PI}{[\ion{P}{i}]}
\newcommand{\CIp}{\ion{C}{i}}
\newcommand{\HeI}{\ion{He}{i}}
\newcommand{\MgIp}{\ion{Mg}{i}}
\newcommand{\MgIIp}{\ion{Mg}{ii}}
\newcommand{\NaI}{\ion{Na}{i}}
\newcommand{\HI}{\ion{H}{i}}
\newcommand{\brg}{Br$\gamma$}
\newcommand{\pab}{Pa$\beta$}

\newcommand{\macc}{$\dot{M}_{acc}$}
\newcommand{\lacc}{L$_{acc}$}
\newcommand{\lbol}{L$_{bol}$}
\newcommand{\mjet}{$\dot{M}_{out}$}
\newcommand{\mh}{$\dot{M}_{H_2}$}
\newcommand{\Ne}{n$_e$}
\newcommand{\h}{H$_2$}
\newcommand{\kms}{km\,s$^{-1}$}
\newcommand{\um}{$\mu$m}
\newcommand{\lam}{$\lambda$}
\newcommand{\msyr}{M$_{\odot}$\,yr$^{-1}$}
\newcommand{\Av}{A$_V$}
\newcommand{\msun}{M$_{\odot}$}
\newcommand{\lsun}{L$_{\odot}$}
\newcommand{\cm}{cm$^{-3}$}

\newcommand{\bet}{$\beta$}
\newcommand{\alfa}{$\alpha$}

\hyphenation{a-na-ly-sis mo-le-cu-lar pre-vious e-vi-den-ce dif-fe-rent pa-ra-me-ters ex-ten-ding a-vai-la-ble ca-li-bra-tion va-ri-a-bi-li-ty}

\title{AMBER/VLTI high spectral resolution observations of the Br$\gamma$ emitting region in HD\,98922 }
\subtitle{A compact disc wind launched from the inner disc region
\thanks{Based on observations collected at the VLT (ESO Paranal, Chile) with programmes 075.C-0637(A), 083.C-0236(A-D), 090.C-0192(A), 090.C-0378(A) and 090.C-0371(A)}}
\author{A. Caratti o Garatti \inst{1,2}, L.V. Tambovtseva \inst{2,3}, R. Garcia Lopez \inst{1,2}, S. Kraus \inst{4}, D. Schertl \inst{2},
V.P. Grinin \inst{2,3,5}, G. Weigelt \inst{2}, K.-H. Hofmann \inst{2}, F. Massi \inst{6},
S. Lagarde \inst{7}, M. Vannier \inst{7}, and F. Malbet \inst{8}}

\offprints{A. Caratti o Garatti, \email{alessio@cp.dias.ie}}

\institute{
Dublin Institute for Advanced Studies, School of Cosmic Physics, Astronomy \& Astrophysics Section,
31 Fitzwilliam Place, Dublin 2, Ireland \\ 
\email{alessio@cp.dias.ie; rgarcia@cp.dias.ie}\\
\and
Max-Planck-Institut f\"{u}r Radioastronomie, Auf dem H\"{u}gel 69, D-53121 Bonn, Germany\\
\email{
ds@mpifr-bonn.mpg.de; gweigelt@mpifr-bonn.mpg.de; khh@mpifr-bonn.mpg.de}\\
\and
Pulkovo Astronomical Observatory of the Russian Academy of Sciences, Pulkovskoe shosse 65, 196140 St. Petersburg, Russia\\ 
\email{lvtamb@mail.ru; grinin@gao.spb.ru}\\
\and
University of Exeter, School of Physics, Stocker Road, Exeter, EX4 4QL, UK
\email{skraus@astro.ex.ac.uk} \\
\and
The V.V. Sobolev Astronomical Institute of the St. Petersburg University, Petrodvorets, 198904 St. Petersburg, Russia\\ 
\and
INAF – Osservatorio Astrofisico di Arcetri, Largo E. Fermi 5, 50125 Firenze, Italy \\
\email{fmassi@arcetri.astro.it}\\
\and
Laboratoire Lagrange, UMR 7293, Universit\'e de Nice Sophia-Antipolis, CNRS, Observatoire de la C\^ote D’Azur, BP 4229, 06304, Nice Cedex 4, France\\
\email{Stephane.Lagarde@oca.eu; martin.vannier@unice.fr}\\
\and
UJF-Grenoble 1/CNRS-INSU, Institut de Plan\'etologie et d'Astrophysique de Grenoble (IPAG) UMR 5274, 38041 Grenoble, France\\
\email{Fabien.Malbet@obs.ujf-grenoble.fr}
}

%
\date{Received date; Accepted date}
%
%
%
\abstract
   {High angular and spectral resolution observations can provide us with fundamental clues to the
   complex circumstellar structure of young stellar objects (YSOs) and to the physical processes taking place close to these sources.}
   {We analyse the main physical parameters and the circumstellar environment of the young Herbig Be star HD 98922.}
   {We present AMBER/VLTI high spectral resolution ($ \mathrm R $=12\,000) interferometric observations across the Br$\gamma$ line, accompanied by
   UVES high-resolution spectroscopy and SINFONI-AO assisted near-infrared (NIR) integral field spectroscopic data.
   To interpret our observations, we develop a magneto-centrifugally driven disc-wind model.}
 {Our analysis of the UVES spectrum shows that HD 98922 is a young ($\sim$5$\times$10$^5$\,yr) Herbig Be star (SpT=B9V), located at a distance of 440$\pm^{60}_{50}$\,pc, 
with a mass accretion rate ($\dot{M}_{acc}$) of $\sim$(9$\pm$3)$\times$10$^{-7}$\,M$_{\sun}$\,yr$^{-1}$.
SINFONI $K$-band AO-assisted imaging shows a spatially resolved circumstellar disc-like region ($\sim$140\,AU in diameter) with asymmetric brightness distribution. 
Our AMBER/VLTI UT observations indicate that the Br$\gamma$ emitting region (ring-fit radius $\sim$0.31$\pm$0.04\,AU) is smaller
than the continuum emitting region (inner dust radius $\sim$0.7$\pm$0.2\,AU), showing significant non-zero V-shaped differential phases (i.e. non S-shaped, as expected for a rotating disc). 
The value of the continuum-corrected pure Br$\gamma$ line visibility at the longest baseline (89\,m)
is $\sim$0.8$\pm$0.1, i.e.  the Br$\gamma$ emitting region is partially resolved. Our modelling suggests that the observed Br$\gamma$ line-emitting region mainly originates from a disc wind with a half opening angle of
30$\degr$, and with a mass-loss rate ($\dot{M}_w$) of $\sim$2$\times$10$^{-7}$\,M$_\sun$\,yr$^{-1}$.
The observed V-shaped differential phases are reliably reproduced by combining a simple asymmetric continuum disc model with our Br$\gamma$ disc-wind model.}
{In conclusion, the Br$\gamma$ emission of HD\,98922 can be modelled with a disc wind that is able to approximately reproduce all interferometric observations
if we assume that the intensity distribution of the dust continuum disc is asymmetric.}

\titlerunning{AMBER/VLTI HR observations of the Br$\gamma$ emitting region in HD\,98922}
\authorrunning{A. Caratti o Garatti et al.}
\maketitle

%

\section{Introduction}
\label{introduction:sec}

Herbig Ae/Be stars are intermediate-mass pre-main-sequence stars~\citep[2--10\,M$_\sun$; see e.g.][]{herbig60,waters98}.
They are the intermediate-mass counterparts of Classical T Tauri stars (CTTSs) and are a fundamental link between low- and high-mass star formation.
As is the case for their low-mass counterparts, the circumstellar environment of Herbig Ae/Be stars is not fully understood.
In particular, their innermost disc regions, within 1\,AU from the central sources, harbour both accretion and ejection processes,
whose study is of fundamental relevance to understanding how stars form.
Herbig Ae/Be stars display several features indicative of their complex circumstellar environments and 
of the accretion and ejection processes taking place close to the sources: infrared excess indicating circumstellar material and disc,
UV excess and veiling~\citep[]{donehew11} indicative of accretion, as well as several permitted and forbidden lines which can
be used to trace both accretion and ejection processes (e.g. \ion{H}{i}, \ion{He}{i}, or [\ion{O}{i}], [\ion{Fe}{ii}] lines).

However, owing to the small spatial scales involved (1\,AU corresponds to 10 milliarseconds -mas- at 100\,pc) high angular resolution is required
to spatially resolve the inner disc region.
Additionally, to study the gas kinematics and disentangle the different processes, high spectral resolution is needed as well. 
Near-IR interferometry at medium and high spectral resolution has thus become a fundamental tool in probing the accretion/ejection processes in 
the inner disc regions~\citep[see e.g.][]{eisner_herbig07,kraus08,eisner09,eisner14,weigelt}. 

Observations have been mostly focused on the Br$\gamma$ line (2.166\,$\mu$m), which is bright in Herbig Ae/Be stars and can  often be associated with accretion or ejection
processes. Notably, only five Herbig Ae/Be stars have been  observed so far through high spectral resolution NIR interferometry~\citep[$ \mathrm R \sim$12\,000; see][]{weigelt,kraus12,garcia13,rebeca15,ellerbroek}.
In particular, VLTI/AMBER observations of the Br$\gamma$ at high spectral resolution have been proved effective in
resolving the disc-wind region and in inferring the flux contribution of the unresolved magnetosphere or X-wind region to the total line flux. 
Our group has successfully developed disc-wind, X-wind, and magnetospheric models that match these observations very well \citep[see][]{grinin11,weigelt,tambovtseva,rebeca15}.
In this paper we  study the circumstellar environment of \object{HD 98922}, through VLTI/AMBER interferometry, along with UVES
(high-resolution UVB spectroscopy) and SINFONI-AO assisted (NIR integral field spectroscopy) ancillary data.

\object{HD 98922} is a Herbig Be star of spectral-type B9V~\citep{houk}, although 
a more recent analysis suggests it might be a later spectral type~\citep[A2III;][]{hales}. 
Its SED has a large infrared excess, 
indicating the presence of an extended dusty disc~\citep{malfait}. Several optical and near-IR emission lines prove its strong circumstellar activity. 
In particular, from the analysis of the [\ion{O}{i}] emission line profiles, \citet{acke05} conclude that
this line cannot originate from the self-shadowed dusty disc surface, but likely from a rotating gaseous disc inside the dust-sublimation radius.
From the analysis of the Br$\gamma$ line, \citet{rebeca06} derive a high mass-accretion rate 
($\dot{M}_{acc} \sim 2 \times 10^{-6}$\,M$_{\sun}$\,yr$^{-1}$, assuming a distance of 540\,pc),
indicating that the Herbig Be star is still actively accreting. AMBER-LR and MR interferometric studies from \citet{kraus08} show that the Br$\gamma$ line visibility 
increases with respect to the continuum visibility. In particular, the Br$\gamma$ line-emitting region is not resolved with baseline
lengths up to $\sim$60\,m (an upper limit of 1\,mas, or 0.5\,AU at a distance of 540\,pc, is reported), whereas the continuum-emitting region ring-fit diameter is estimated 
to be 4.6$\pm$0.1\,mas (or 2.48$\pm$0.05\,AU at a distance of 540\,pc). 
The authors conclude that the Br$\gamma$ line could originate from a stellar wind, X-wind, 
or magnetospheric accretion, and they favour the last scenario.
Moreover, a strong wind/outflow activity is also observed from the P\,Cygni profiles of the H$\alpha$ and \ion{Si}{ii} lines~\citep[$v_{rad}\sim$ 300\,km\,s$^{-1}$;][]{grady} 
at optical wavelengths, as well as from the \ion{He}{i} line profile~\citep{oudmaijer11} in the near-IR. 

Visual extinction values towards the object are between 0.3 and 0.5\,mag, and the average magnitude in the V band is 6.67\,mag~\citep[see e.g.][]{manoj}. 
The inclination of the \object{HD 98922} system axis with respect to the plane of the sky is poorly known. A very rough estimate of 45$\degr$ was derived in \citet{blondel} by modelling 
the UV spectrum, whereas \citet{hales} report an estimate of 20$\degr$ from CO modelling.

Finally, the distance to the source is quite uncertain. Trigonometric distance measured with the HIPPARCOS satellite~\citep{vandenAncker98} gives 
a value of $d>$540\,pc, which is well beyond the HIPPARCOS observational range (d $\sim$ 350--400\,pc). A later revision of these data provided a larger value 
of $d=1190^{+930}_{-390}$~\citep{vanLeeuwen}. This  value translates into an extremely large stellar luminosity for \object{HD 98922} of $\sim$10$^4$\,L$_{\sun}$. 
As a consequence its position in the 
Hertzsprung–Russell diagram is well outside the locus of the pre-main sequence Herbig Ae/Be stars~\citep[see e.g. Fig.\,4 in][see also a more detailed discussion 
in Sect.~\ref{physpars:sec}]{alecian13}, namely its position cannot be reproduced by using the evolutionary tracks of Herbig Ae/Be stars.
Recently, the analysis of \citet{hales} has provided a closer distance of 507$^{+131}_{-104}$\,pc.
It is also worth noting that this source is a wide binary~\citep[$\sim$8$\arcsec$, PA\,=\,343$\degr$][]{dommanget} and it is possible that  
a second closer companion might also be present, as hinted by spectro-astrometric observations~\citep[$<$0\farcs5, PA$\sim$0$\degr$][]{baines}. 

The paper is structured as follows. Spectroscopic and interferometric observations and data reduction are presented in Sect.~\ref{observations:sec}. 
Section~\ref{spec-results:sec} reports results from UVES high-resolution spectroscopy, along with the derived stellar physical parameters
and distance. Section~\ref{SINFONI-results:sec} presents the results of our SINFONI observations.
We present our interferometric results across the Br\,$\gamma$ line, along with its 
geometrical and physical modelling in Sects.~\ref{AMBER_results:sec} and \ref{model:sec}.
Section~\ref{discussion:sec} provides a discussion on the origin of the Br\,$\gamma$ emitting region and of the asymmetries observed in the interferometric observables.
Finally, our conclusions are presented in Sect.~\ref{conclusion:sec}.

\section{Observations and data reduction}
\label{observations:sec}

\subsection{VLT/UVES spectroscopy}

HD 98922 was observed on  21  March 2005 with UVES~\citep{dekker}, the echelle spectrograph mounted on the VLT/UT2. 
The observations were performed with the blue arm, covering a spectral range between $\sim$3750\,\AA 
~and $\sim$4990\,\AA. A slit width of 0\farcs4, which
gives a spectral resolution of  80\,000, was adopted, and the total integration time was 100\,s. 
Wavelength calibration frames were taken with a long slit and a ThAr arc lamp. The data were reduced using the UVES pipeline v3.2.0~\citep{ballester} 
available from the ESO Common Pipeline Library. We converted the wavelength scale to the heliocentric rest frame. 
No standard star is available to provide an accurate flux calibration. Owing to the very small photometric variability of HD 98922~\citep{dewinter}, 
we adopt the photometric value reported in \citet{manoj} ($m_B$=6.81\,mag) to calibrate the spectrum.

\subsection{SINFONI AO spectral-imaging}

HD 98922 was observed on  2  February 2013 with the VLT/SINFONI integral field spectrograph~\citep[][]{sinfoni} in the
K-band (1.95--2.45\,$\mu$m) at the highest spectral ($ \mathrm R \sim$4000) and spatial resolution
(12.5$\times$25\,mas pixel scale).  
The detector integration time (DIT) and number of sub-exposures per frame were 2\,s and five, respectively, both for the on-source (target) and the off-source (sky) position.
In total, eight target-sky pairs were obtained at position angles (PAs) 0$\degr$, 90$\degr$, 180$\degr$, and 270$\degr$ (two pairs per PA), corresponding to a total integration
time on-source of 80\,s.
These observations were AO-assisted (using HD 98922 as a natural guide star), covering a field
of view (FoV) of 0.8$\arcsec \times$0.8$\arcsec$.

To correct for atmospheric transmission, observations of a telluric standard star of spectral type A
were also performed. 
The main data reduction process was done using the SINFONI data-reduction pipeline in GASGANO~\citep[][]{modigliani}, i.e. dark and bad
pixel masks, flat-field corrections, optical distortion correction, and wavelength calibration using arc lamps.
As a result, we obtained four calibrated datacubes, one for each observed PA.

We then measured the full width at half maximum (FWHM) of both target and standard star on the final datacubes to estimate the achieved spatial resolution,
which is $\sim$60\,mas. 
From each datacube the stellar spectrum was then extracted, and continuum and Br$\gamma$ line images
were created by collapsing the datacubes along the z-axis, from 2.15 to 2.16\,$\mu$m and from 2.164 to 2.168\,$\mu$m, respectively.

\begin{table*}[t]
 \centering
\caption{\label{tab:obs} Log of the VLTI/AMBER/FINITO observations of HD~98922 and calibrators.}
\begin{scriptsize}
\begin{tabular}{ccccccccccc}
\hline\hline
HD 98922    & \multicolumn{2}{c}{Time [UT]}& Unit Telescope & Spectral & Wavelength  & DIT\tablefootmark{a} & N\tablefootmark{b} & Seeing  & Baselines & PAs  \\
Observation  & Start & End                  & array          & mode\tablefootmark{c} & range       &        &   &       &          &      \\
Date         &       &                      &                &          & ($\mu$m)    & (s)    &   &($\arcsec$) &    (m)   & ($\degr$)  \\
\noalign{\smallskip}
\hline
\noalign{\smallskip}
2012 Dec 26 & 09:02 & 09:20                & UT2-UT3-UT4    &  HR-K-F  & 2.147-2.194 & 1       & 100    & 0.7-0.8 & 44.1/60.6/89.4 & 219.3/103/76.7    \\
2013 Feb 28 & 04:50 & 05:32                & UT2-UT3-UT4    &  HR-K-F  & 2.147-2.194 & 1       & 1540     & 0.9-1.2 & 43.9/61/89.4 & 220.9/104.9/78.8  \\
\noalign{\smallskip}
\hline
\end{tabular}

\vspace*{3mm}
\begin{tabular}{ccccccccccc}
\hline\hline
Calibrator    & Date         & \multicolumn{2}{c}{Time [UT]} & UT array    & Spectral & Wavelength  & DIT\tablefootmark{a} & N\tablefootmark{b} & Seeing  & Uniform-disc \\
Name          &              & Start & End                   &             & mode\tablefootmark{c} & range       &         &       &         & diameter\tablefootmark{d} \\
       &              &       &                       &             &          & ($\mu$m)    & (s)     &       & ($\arcsec$)& (mas) \\
\noalign{\smallskip} \hline \noalign{\smallskip}
HD 60276  & 2012 Dec 26 & 07:50 & 08:03                 & UT2-UT3-UT4 &  HR-K-F  & 2.147-2.194 & 1       & 200   & 0.9-1.0 & 0.92$\pm$0.06 \\
HD 103125 & 2013 Feb 28 & 04:08 & 04:42                 & UT2-UT3-UT4 &  HR-K-F  & 2.147-2.194 & 1       & 70    & 0.9-1.5 & 0.88$\pm$0.06 \\

\noalign{\smallskip} \hline
\end{tabular}
\end{scriptsize}
\begin{flushleft}
\hspace{0mm}
\tablefoot{
\tablefoottext{a}{Detector integration time per interferogram.}
\tablefoottext{b}{Number of interferograms.}
\tablefoottext{c}{High spectral resolution mode in the $K$-band using the fringe tracker FINITO.}
\tablefoottext{d}{UD diameter taken from JMMC Stellar Diameters Catalogue - JSDC \citep{lafrasse10}.}
}
\end{flushleft}
\end{table*}

\subsection{VLTI/AMBER/FINITO Interferometry}

The log of our interferometric observations is reported in Table\,\ref{tab:obs}.
\object{HD 98922} was observed with AMBER~\citep[][]{petrov07}, the NIR beam-combiner of
the Very Large Telescope Interferometer (VLTI) operated by ESO, during two different runs on  26 December 2012 and 28  February 2013.
On both occasions we employed AMBER’s high spectral resolution mode in the $K$-band (HR mode; $ \mathrm R $ = 12\,000 or $\Delta$v $\sim$ 25\,km\,s$^{-1}$) 
covering the spectral range from 2.147 to 2.194\,$\mu$m, centred on the Br$\gamma$ line emission (at 2.166\,$\mu$m).
Observations were conducted with the UT2-UT3-UT4 telescopes, using the fringe tracker FINITO~\citep[][]{gai} 
for co-phasing and a detector integration time of 1.0\,s per interferogram. In the first run 100 interferograms were recorded, whereas
$\sim$1500 were recorded in our second run. Stars HD 60276 and HD 103125 were observed with the same observational settings
and were used as interferometric calibrators to derive the transfer function for the first and second run, respectively.
The calibration datasets for the first and second runs consist of 200 and 980 calibrator interferograms, respectively.

To reduce our interferograms, we used our own data reduction software based on the P2VM algorithm~\citep[][]{tatulli07}, which
provides us with wavelength-dependent visibilities, wavelength-differential phases, closure phases, and wavelength calibrated spectra.

The fringe tracking performance of FINITO typically differs during the observations of target and calibrator.
Therefore, we used the archival low-resolution visibilities of HD\,98922 (216 different visibility {\it uv} points, mostly taken during the AMBER's GTO)
to calibrate the continuum visibilities at the six $uv$ points of our HR observations.
This calibration method provides us with a visibility error of about $\pm$5\% due to the large number of employed 
LR visibility measurements.
By using low-resolution spectroscopic data taken at different epochs to calibrate our data, we are implicitly assuming that 
both size and morphology of the object have not changed more than  the data uncertainties allow.

The wavelength calibration of the AMBER data was refined using the numerous telluric lines present in the 2.15--2.19\,$\mu$m
region~\citep[see][for more details on the wavelength calibration method]{weigelt}. 
We estimate an uncertainty in the wavelength calibration of $\sim$0.2\,\AA~or $\sim$3\,km\,s$^{-1}$.

\section{Results from UVES spectroscopy}
\label{spec-results:sec}

\begin{figure}
\centering
\includegraphics [width=9.1cm]{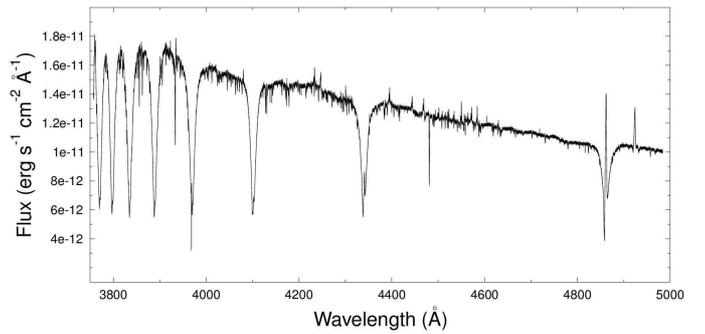}\\
\caption{UVES/VLT flux-calibrated spectrum of HD\,98922.}
\label{fig:UVESspectrum}
\end{figure}

\subsection{Line recognition}
\label{lines:sec}

\begin{figure*}
\resizebox{\textwidth}{!}{\includegraphics{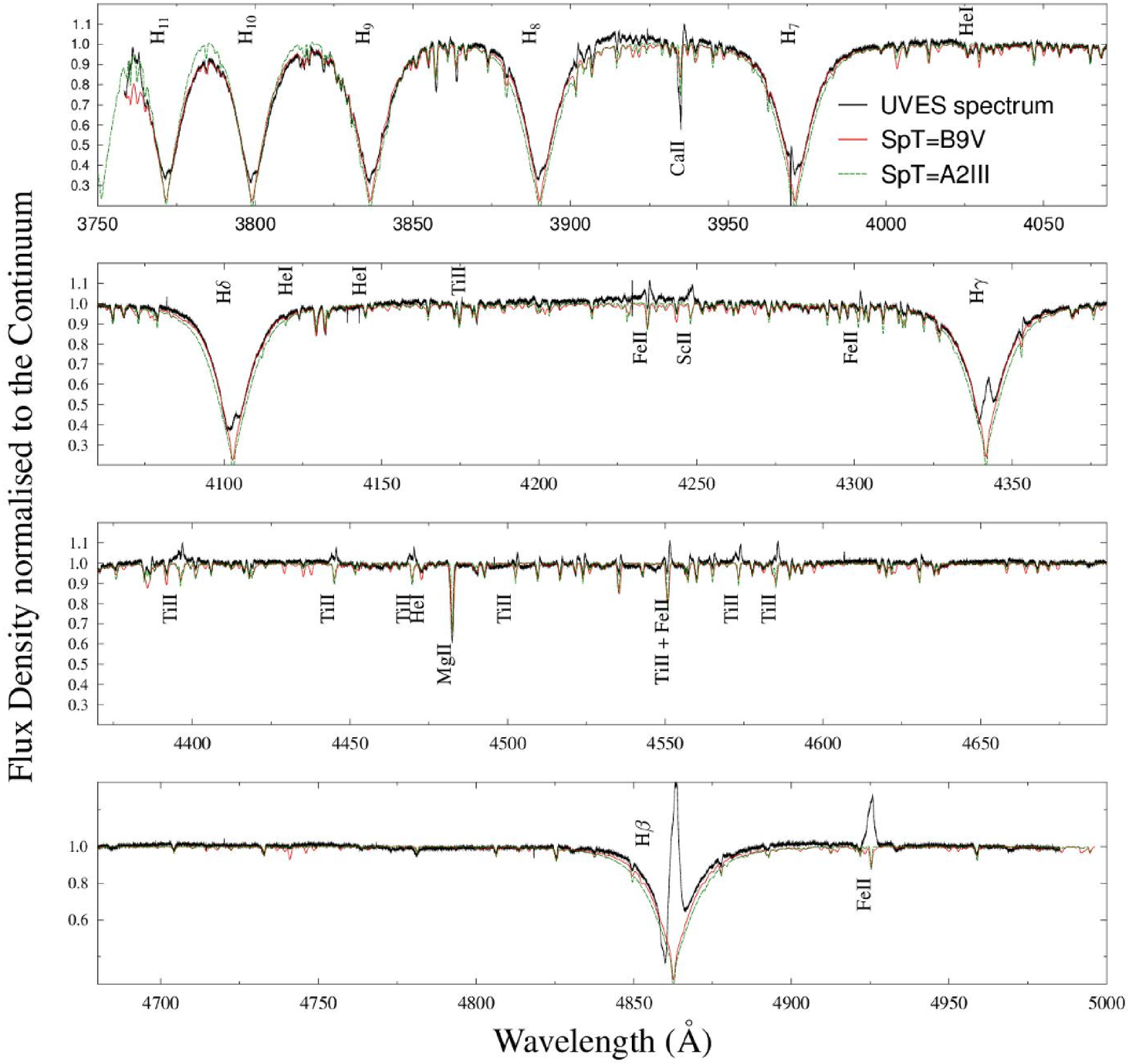}}
\caption{UVES/VLT high-resolution spectrum of HD\,98922 normalised to the continuum (black solid line). 
The red solid line shows the best-fitting synthetic spectrum, $SpT$=B9V, $log\,g$=3.5, $[Fe/H]$=-0.5.
For comparison, a synthetic spectrum with parameters $SpT$=A2III, $log\,g$=3.0, $[Fe/H]$=-0.5 (green dashed line) is also shown.
The most prominent spectral features are  labelled.
}
\label{fig:normspectrum}
\end{figure*}

The UVES flux-calibrated spectrum of HD\,98922 is presented in Figure~\ref{fig:UVESspectrum} and  shows the descending stellar continuum with several strong 
absorption lines and few faint emission features. 
The same spectrum, normalised to the continuum, is shown in Figure~\ref{fig:normspectrum} (black line).
The most prominent features, both in absorption and emission, are also labelled in Fig.~\ref{fig:normspectrum}.

The most prominent absorption features are the \ion{H}{i} lines from the Balmer series (from \ion{H}{11} to H$\beta$), 
which partially show emission due to the circumstellar environment.
These lines show P\,Cygni profiles (see e.g. the H$\beta$ line vs. its theoretical profile in Fig.~\ref{fig:normspectrum}: the peak in emission is blue-shifted with respect to the zero velocity, and
the blue-shifted absorption is deeper than the theoretical value), indicative of winds or outflows.
The remaining absorption features are mostly photospheric from atomic lines like \ion{He}{I}, \ion{Ca}{II}, \ion{Fe}{I}, \ion{Fe}{II},
\ion{Mg}{I}, \ion{Mg}{II}, \ion{Si}{I}, and \ion{Si}{II}.
Other emission features, mostly \ion{Fe}{II} and \ion{Ti}{II} lines, located between $\sim$4200 and 4600~\AA, arise from circumstellar activity.
These lines are characterised by a double-peaked profile, are likely produced by self-absorption of the photospheric feature, and have a broad pedestal
(full width zero intensity, $FWZI\sim$4--4.4~\AA, i.e. $\Delta \mathrm{v} \sim$270--290\,km\,s$^{-1}$), which originates in a region with a large radial velocity structure, 
likely located in the magnetospheric accretion funnels or in the hot inner region of the circumstellar disc~\citep{beristain,aspin}. 
The strongest \ion{Fe}{ii} emission line (at 4925~\AA) shows a P\,Cygni profile.

From the analysis of the \ion{H}{11} line, we derive a value of $\sim$-8\,km\,s$^{-1}$ for the stellar radial velocity ($\mathrm{v}_{rad}$) with respect to the local standard of rest.
Notably, the observed spectral lines do not show any indication of binarity: no splitting and no additional absorption lines are detected in the spectrum.

\subsubsection{Stellar physical parameters and distance}
\label{physpars:sec}

To allow both identification and analysis of the observed line profiles we used synthetic spectra from the PHOENIX library~\citep{husser}, which 
provides high-resolution ($ \mathrm R $=500\,000) synthetic spectra based on the stellar atmosphere code PHOENIX. The spectral resolution 
was then reduced to that of the  UVES observations by convolving the synthetic spectra with a Gaussian profile (IRAF task {\em gauss}). 
We constructed a grid of spectra and varied different stellar parameters, namely $T_{eff}$ (from 9000 to 11\,000\,K, $\Delta T$ = 200\,K), 
log\,$g$ (from 3.0 to 4.5, $\Delta$log\,$g$ = 0.5), and [Fe/H] (from -1 to 1, $\Delta$[Fe/H] = 0.5). 
Both observed and synthetic spectra were then normalised to the continuum. 
Our own IDL programme was used to match the UVES spectrum with the best-fitting model in three steps. 
First, by comparing the wings of the Balmer lines the stellar gravity was inferred. 
In this case, Balmer lines from H11 to H$\gamma$ were used (the H$\beta$ line was excluded because it is too  affected by circumstellar emission).
Second, the effective temperature ($T_{\mathrm{eff}}$) of the star was obtained by comparing observed and theoretical equivalent widths (EWs)
of lines sensitive to $T_{\mathrm{eff}}$, namely the \ion{He}{i} lines (at 4026, 4120, 4145, 4470~\AA), and the \ion{Ti}{ii} line at 4176~\AA.
Finally, the metal abundance was inferred by comparing the iron abundances for a well-defined set of \ion{Fe}{I} and \ion{Fe}{II} lines.
As a result, we obtained the stellar effective temperature, gravity, and metallicity, namely 
$T_{\mathrm{eff}}$ = 10\,400$\pm$200\,K (i.e. spectral type B9V), $log\,g$ = 3.5$\pm$0.2, and $[Fe/H]$ = -0.5$\pm$0.2. The best-fitting synthetic spectrum, normalised to the
stellar continuum, is shown in Figure~\ref{fig:normspectrum} (red solid line) superimposed on the observed spectrum (in black).
For comparison, Fig.~\ref{fig:normspectrum} also shows a normalised synthetic spectrum (green dashed  line) with the stellar parameters derived by \citet{hales}, namely 
$T_{\mathrm{eff}}$ = 9000\,K (SpT=A2 III), $log\,g$=3.0, and $[Fe/H]$=-0.5.

To derive the remaining stellar parameters, we follow \citet{montesinos}. We place the estimated values of the stellar parameters on a log\,$T_{\mathrm{eff}}$, log\,$g$ HR diagram, 
using evolutionary tracks and isochrones from \citet{siess00} with $[Fe/H]$=-0.5. This provides us
with an estimate of the stellar mass ($M_*$=5.2$\pm$0.2\,M$_{\sun}$) and age ($\sim$5$\times$10$^5$\,yr) for HD\,98922, and, in turn,
stellar luminosity ($L_*$ = 640$\pm^{130}_{160}$\,L$_{\sun}$), radius ($R_*$ = 7.6$\pm^{0.6}_{2}$\,R$_{\sun}$), and visual absolute magnitude
(M$_V$=-1.75$\pm$0.3\,mag) can be inferred from a (log\,$T_{\mathrm{eff}}$ $vs.$ log\,$L_*$) HR diagram.

Following \citet{montesinos}, we also aim at deriving a better estimate of the distance ($d$) to HD\,98922, from the distance module equation:

\begin{equation}
\label{m-M:eq}
d=10^{(m_V - A_V - M_V +5)/5}.
\end{equation}We assume m$_V$=6.77\,mag and A$_V$=0.3~\citep{rebeca06}, and we obtain $d$=440$\pm^{60}_{50}$\,pc. By adopting this distance value, an estimate of the mass accretion rate ($\dot{M}_{acc}$)
can be also inferred from the EW of the Br$\gamma$ line after correcting for the visual extinction and self-absorption
from the photosphere~\citep[see e.g.][]{rebeca06}. By using their published data we obtain
$\dot{M}_{acc}$$\sim$(9$\pm$3)$\times$10$^{-7}$\,M$_{\sun}$\,yr$^{-1}$.

All stellar and disc parameters used in this paper, derived in this work or adopted from the literature, are listed in Table~\ref{stellar_parameters:tab}.

\begin{table}
 \caption[]{\label{stellar_parameters:tab} HD\,98922 stellar and disc parameters}
\begin{tabular}{lcc}
 \hline \hline\\[-5pt]
Stellar Parameter &  Value & Reference \\
 \hline\\[-5pt]
Distance   & 440$\pm^{60}_{50}$\,pc & 1\\
$T_{\mathrm{eff}}$ &  10\,400$\pm$200\,K & 1,2\\
$Sp Type$ & B9V  & 1,2 \\
$log\,g$  & 3.5$\pm$0.2 &  1\\
$[Fe/H]$ & -0.5$\pm$0.2 & 1 \\
$M_*$   & 5.2$\pm$0.2\,M$_{\sun}$ & 1\\
$R_*$   & 7.6$\pm^{0.6}_{2}$\,R$_{\sun}$ & 1\\
$L_*$  & 640$\pm^{130}_{160}$\,L$_{\sun}$ & 1\\
age & 5$\times$10$^5$\,yr & 1\\
M$_V$ & -1.75$\pm$0.3\,mag  & 1 \\
m$_V$ & 6.77\,mag & 3 \\
A$_V$ & 0.3  & 3 \\
m$_K$ & 4.28 & 4 \\
F$_{obs}(K)$/F$_{*}(K)$ & 4.3 & 1\\
$i$ &  20$\degr$ & 5\\
$\mathrm{v}_{rot}\,sin\,i$ & 50$\pm$3\,km\,s$^{-1}$  &  6  \\
$\dot{M}_{acc}$ & (9$\pm$3)$\times$10$^{-7}$\,M$_{\sun}$\,yr$^{-1}$ & 4 + 1 \\
\hline
\end{tabular}
\tablebib{(1)~This work; (2)~\citet{the}; (3)~\citet{manoj}; (4)~\citet{rebeca06}; (5)~\citet{hales}; (6)~\citet{alecian13};
}
\end{table}

\section{Results from SINFONI-AO spectral imaging}
\label{SINFONI-results:sec}

\begin{figure}[h!]
\includegraphics[width=8.5cm,trim= 0 20 0 0]{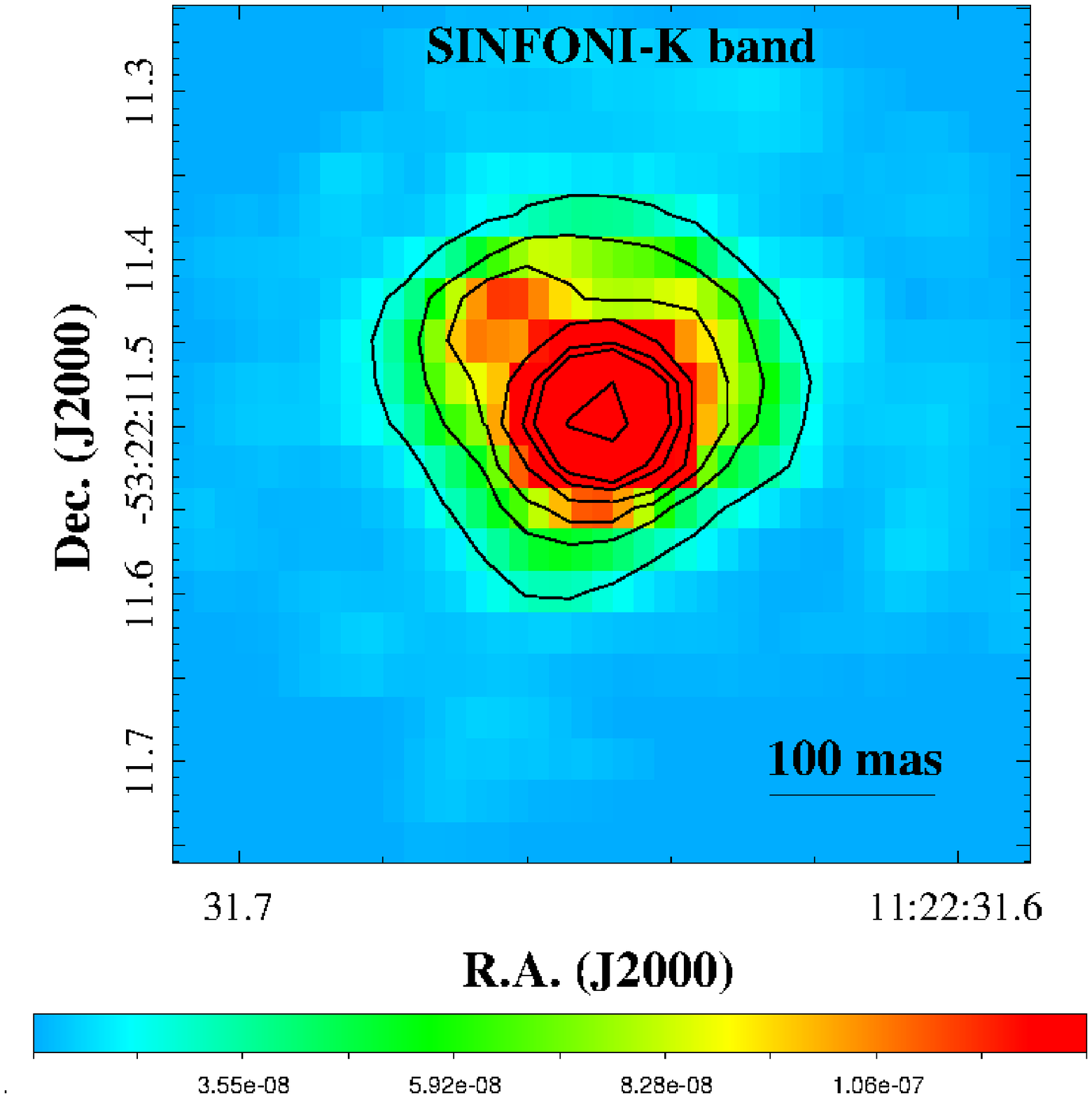}
\includegraphics[width=8.5cm,trim= -31 -90 15 10]{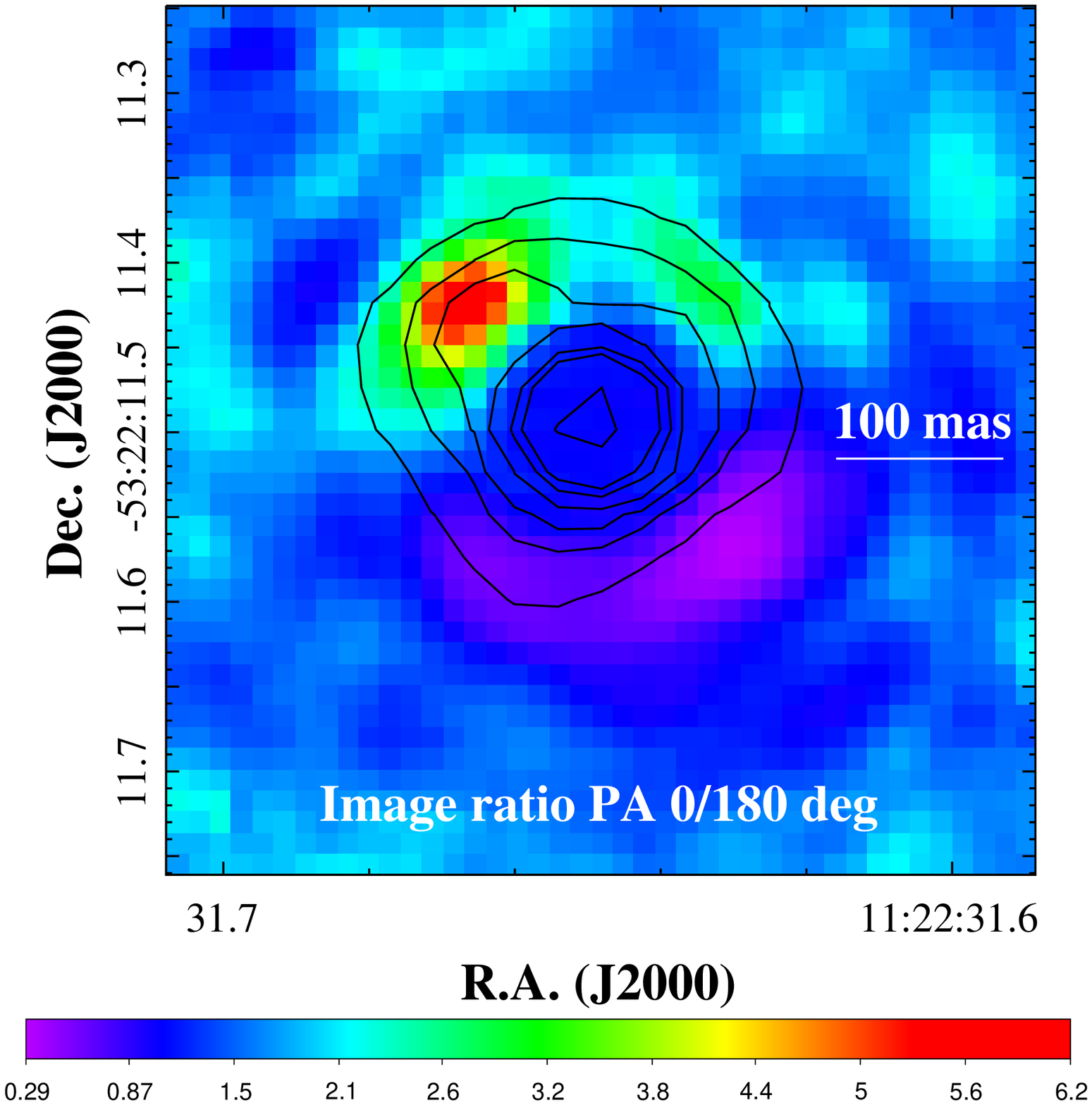}
\caption{\label{fig:disc_asymmetry} 
\textit{Upper Panel}: SINFONI K-band continuum image (between 2.15 and 2.16\,$\mu$m).
Labelled fluxes are in W\,m$^{-2}$\,$\mu$m\,arcsec$^{-2}$.
Contour levels are 4, 5, 50, 150, 250, 400, 500, and 1000\,$\sigma$.
\textit{Lower Panel}: Intensity ratio image of two SINFONI K-band continuum images (between 2.15 and 2.16\,$\mu$m)
taken at PA of 0$\degr$ and 180$\degr$, showing the asymmetric disc emission. Contours show
the K-band continuum of the PA=0$\degr$ image (see upper panel), and give the position of both symmetric PSF in the centre,
and asymmetric disc emission. Contour levels are as in the continuum image.
}
\end{figure}

Our SINFONI K-band spectral images resolve a bright asymmetric scattered emission (diameter $\sim$320\,mas at 3$\sigma$, or
$\sim$140\,AU at a distance of 440\,pc) that extends beyond the symmetric point spread function (PSF; FWHM=60\,mas) of HD\,98922 
(see the upper panel of Figure~\ref{fig:disc_asymmetry} and its contour levels).
This structure is not an artefact and because it is visible in all the spectral images and is co-rotating in each image FoV according to the different observed PAs.

The bright circumstellar structure most likely originates from the scattered light of the disc.
Of particular interest is its asymmetric morphology.
The brightness distribution is roughly represented by a symmetric PSF plus a bright elliptical
arc towards the north-north-east  (NNE) (i.e. a horseshoe-like region), extending from PA$\sim$290$\degr$ to $\sim$110$\degr$ (see upper panel of Fig.~\ref{fig:disc_asymmetry})
and located at $\sim$90\,mas (or $\sim$40\,AU) from the source.
This region is about 2.5 times brighter than its symmetric counterpart located towards south-south-west (SSW).
The value of this ratio was obtained after integrating the flux in two regions (positioned outside the PSF, i.e. 60\,mas), which have the same area and are symmetrical with respect to the central PSF.
It is also worth noting that the brightness distribution of the NNE arc is not homogeneous; it shows an elongated peak between PA$\sim$23$\degr$ and
PA$\sim$65$\degr$, and it is up to three times brighter than the whole arc-shaped feature. The total flux of this area is about one tenth of the flux of the central PSF. 
Because of its elongated arc-shape, which cannot be fitted with the instrumental PSF, we exclude that this structure can be 
the unresolved PSF of a possible companion. 
Remarkably, our SINFONI-AO continuum images do not show any companion between $\sim$800\,mas (or $\sim$350\,AU)
and $\sim$60\,mas (or 26.4\,AU) from the central source. The lower limit of the flux density in our data is $\sim$5$\times$10$^{-11}$\,erg\,s$^{-1}$\,cm$^{-2}$\,$\mu$m$^{-1}$, 
which corresponds to $K \sim 9.8$\,mag or $M_K \sim 1.6$\,mag at the estimated distance of 440\,pc. Adopting the \citet{siess00} evolutionary tracks and
assuming an age of $\sim$5$\times$10$^5$\,yr for any possible undetected companion, the $M_K$ estimate provides us with an upper limit constraint on the spectral type and mass
of SpT$\sim$M0 and $M_*$$\sim$0.5\,M$_\sun$.

To better display the observed brightness asymmetry in the disc, the lower panel of Figure~\ref{fig:disc_asymmetry} shows an intensity ratio image that
is computed from two
SINFONI K-band continuum images (between 2.15 and 2.16\,$\mu$m) with PAs of 0$\degr$ and 180$\degr$, respectively.
The central unresolved PSF disappears (ratio value $\sim$1), whereas 
the bright NNE arc becomes clearly visible (ratio values between 1.5 and 5) in the image.
Apart from the aforementioned elongated peak, there is a dip at PA$\sim$0$\degr$ and a second  arc-shaped peak towards the north-west that is less bright. 

It is worth mentioning that the PIONIER/VLTI reconstructed images of the HD 98922 inner disc~\citep[see Figure\,2 in][]{kluska},
show a similar brightness disc asymmetry, namely the north-eastern side of the disc is brighter than the south-western side.
These interferometric data were acquired between 20 December 2012  and 20 February 2013 (Kluska priv. comm.), simultaneous
with our AMBER and SINFONI observations.
The observed PIONIER disc has, however, a size of $\sim$3-4\,mas ($\sim$1-1.5\,AU), which is well within the (spatially unresolved) PSF.
This suggests that the disc might have a similar morphology from $\sim$1 to 60-70\,AU from the source and that the origin of this brightness 
asymmetry might be the same.

The Br$\gamma$ line is the only emission line detected over the $K$-band continuum of the analysed SINFONI spectral-images.
This emission is not spatially resolved, being enclosed within 60\,mas (or $\sim$26\,AU) from the central source.

\section{Results from VLTI/AMBER interferometry and geometric modelling}
\label{AMBER_results:sec}

\begin{figure}[h!]

        \centering
        \includegraphics[width= 14cm, angle =0]{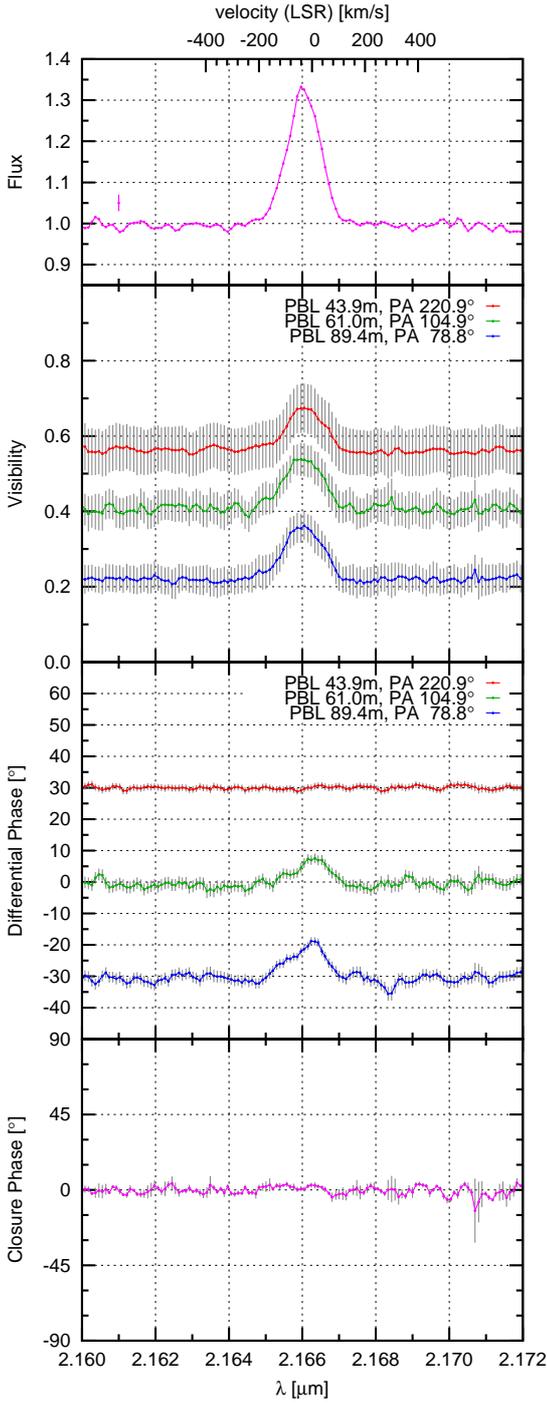}

                \caption{\label{fig:total_v} AMBER observation of HD~98922 at spectral resolution of $ \mathrm R $=12\,000 (28 Feb 2013). 
                From  top to bottom: wavelength dependence of flux, visibilities, wavelength-differential phases, and closure phase observed at three different projected baselines (see labels in figure).
                For clarity, the differential phases of the first and last baselines are shifted by +30$\degr$ and -30$\degr$, respectively. }

\end{figure}

\subsection{Interferometric observables: Visibilities, differential phases, and closure phase}
\label{sect:observables}

Our VLTI/AMBER observations provide us with four direct observables, namely the Br$\gamma$ line profile, visibilities, differential phases, and closure phases. 
These observables allow us to retrieve information about the inner-disc emitting region of HD\,98922, in particular the size and kinematics of the Br$\gamma$ emitting region,
as well as its displacement with respect to the continuum emission.

First of all, we stress that observations in both VLTI runs were taken with the same AMBER settings and UT configurations, which result in similar projected baselines 
and position angles (see Cols.\,10 and 11 of Table\,\ref{tab:obs}).
The results from both runs are identical inside the error bars. However, those from
the second run (28 February 2013) are clearly much less noisy than those from the first run (26 December 2012)
owing to the longer total integration time.
For this reason, in this section we only present the results from the second run (Figure\,\ref{fig:total_v}), whereas
those from the first run are shown in Appendix~\ref{appendix:sec} (Figure\,\ref{fig:total_v2}).
The different panels in Figures\,\ref{fig:total_v} and \ref{fig:total_v2} show (from top to bottom) the Br$\gamma$ line profile (upper panels), visibilities (middle upper panels), 
differential phases (middle lower panels), and closure phases (lower panels).

The wavelength-dependent visibilities (middle upper panels of Fig.~\ref{fig:total_v} and \ref{fig:total_v2}) clearly increase across the Br$\gamma$ line-profile at all baselines. 
This indicates that the Br$\gamma$ emitting region is more compact than the continuum emitting region. 
Previous spectro-interferometric studies of HD\,98922 at medium resolution ($R= 1500$) have also detected an increase in the visibility within the Br$\gamma$ line \citep{kraus08}
with baseline lengths up to 60\,m. 
Our HR mode AMBER observations allow us, for the first time, to measure the visibilities and phases in $\sim$30 different spectral channels across 
the Br$\gamma$ line with baseline lengths up to 90\,m, providing a higher spatial and spectral resolution.
Furthermore, our data indicate that the visibilities from the three baselines are slightly asymmetric with respect to the Br$\gamma$ peak,
namely the visibilities of the blue-shifted wings are systematically smaller than those of the red-shifted wings. 
This might indicate that, at high velocities, the red-shifted Br$\gamma$ emitting region is slighlty more compact than the blue-shifted one, 
or that there is a highly asymmetric continuum.

Differential phases ($\Phi$; lower middle panels of Fig.~\ref{fig:total_v} and \ref{fig:total_v2}) measure the displacement of the line-emitting region
with respect to the continuum-emitting region. For example, the accuracy reachable with a 90\,m baseline and with $\Phi\sim$10$\degr$ is about 0.14\,mas, or 
0.06\,AU at a distance of 440\,pc. Therefore the observed displacements provide us with a very
sensitive measurement of the gas kinematics on scales of a few stellar radii.
For the first time in this object, we detect a clear photocentre shift of the Br$\gamma$ line with respect to its
continuum at the 61\,m and 89\,m baselines. The line displacement is observed in different velocity channels (both blue- and red-shifted). Notably,
differential phases in both blue- and red-shifted channels have the same sign, and
therefore they do not exhibit the typical ``S'' shape usually observed in rotating discs.
Additionally, the red-shifted channels in both baselines show larger differential phases with respect to the blue-shifted ones.
Because these are line-to-continuum phases, the observed shifts in the differential phases could be caused by continuum shifts, line shifts, or both.
The maximum value of the line/continuum displacement is 0.02$\pm$0.04\,mas, 0.14$\pm$0.04\,mas, and 0.16$\pm$0.03\,mas for the 43.9\,m, 61\,m, and 89.4\,m baselines, respectively. 

Finally, it is also worth  noting that the closure phases do not show any significant deviation from zero within the error bars 
(see lower panels of Figs.~\ref{fig:total_v} and \ref{fig:total_v2}); we measure an average value of -0.44$\degr$$\pm$2.55$\degr$.

\subsection{Geometric modelling: size of Br$\gamma$ and continuum emitting region}
\label{sect:line_v}

\begin{figure}[h!]
\centering
{\includegraphics[width= 14cm, angle =0]{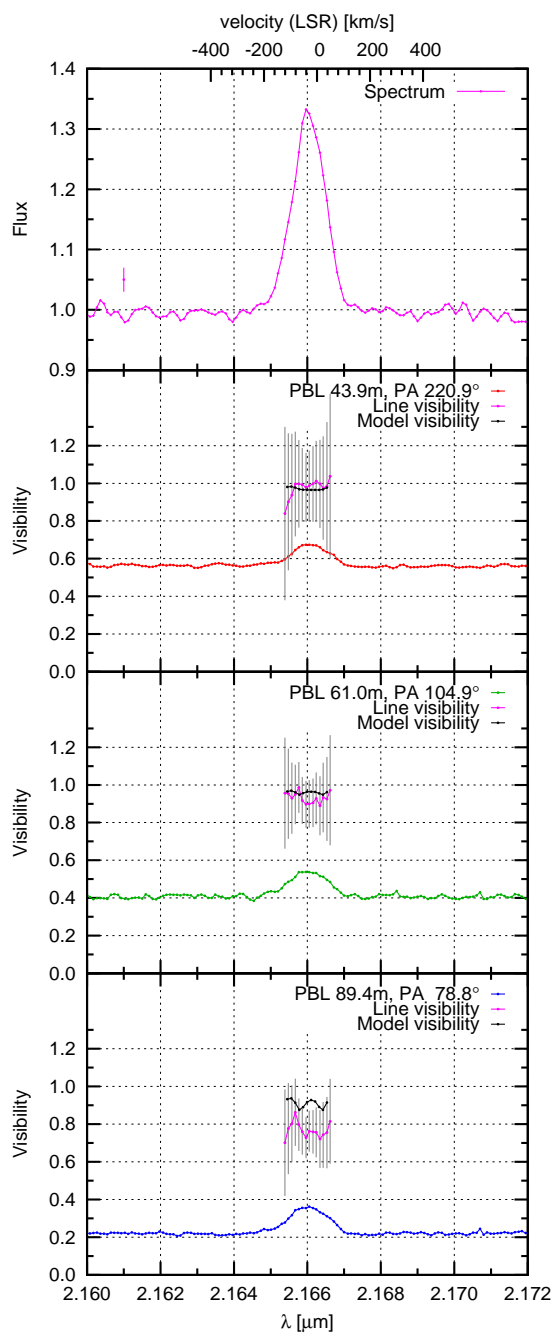}}
        \caption{\label{fig:line_v} Comparison of the observed (pink solid lines) and modelled (black solid lines) continuum-corrected (pure) Br$\gamma$ line visibilities of our AMBER observation of HD~98922.
                From  top to bottom: wavelength dependence of flux and visibilities of the first, second, and third baseline.
                In each visibility panel  the observed total visibilities (red, green, blue, as in Fig.\,\ref{fig:total_v}),  the observed
                pure Br$\gamma$ line visibilities (pink), and the modelled pure Br$\gamma$ line visibilities (black) are shown.
        }
\end{figure}

Our AMBER high spectral resolution data also provide us with a measurement of the visibilities for the pure Br$\gamma$ line-emitting region in the
different spectral channels across the Br$\gamma$ emission line (see Figure~\ref{fig:line_v}). 
The continuum-corrected line visibilities are fundamental for measuring the size of the Br$\gamma$ line-emitting region. 
Within the  wavelength region of the Br$\gamma$ line emission, the measured visibility has two components: the pure
line-emitting part and the continuum-emitting part, which includes continuum emission from both the circumstellar environment and the unresolved
central star. Therefore, following \citet{weigelt07}, the emission line visibility $V_{{\rm Br}\gamma}$ in each spectral channel can be written as
%
%
\begin{eqnarray}\label{eq_brgvis1}
        F_{{\rm Br}\gamma}V_{{\rm Br}\gamma} = \sqrt{ |F_{\rm tot}V_{\rm tot}|^2 + |F_{\rm c}V_{\rm c}|^2 - 2\,F_{\rm tot}V_{\rm tot}\,F_{\rm c}V_{\rm c}\cdot cos{\Phi} }
,\end{eqnarray}
%
where F$_{Br\gamma}$ is the wavelength-dependent line flux, 
$V_{\rm tot}$ ($F_{\rm tot}$) is the measured total visibility (flux) in the  ${\rm Br}\gamma$ line, $V_{\rm c}$ ($F_{\rm c}$) is the 
visibility (flux) in the continuum, and $\Phi$ is the measured wavelength-differential phase within the ${\rm Br}\gamma$ line. 

In our analysis, we also included the intrinsic Br$\gamma$ photospheric absorption feature 
of HD\,98922, using a synthetic spectrum with the spectral type and surface gravity values as in Table~\ref{stellar_parameters:tab}. 

The size of the continuum was obtained using the AMBER GTO archival data at low resolution. 
By fitting a circular symmetric Gaussian model, we derived a Gaussian half width at half maximum (HWHM) radius of 1.8$\pm$0.3\,mas (or $\sim$0.79$\pm$0.13\,AU at a distance of 440\,pc).
A similar value of 1.6$\pm$0.3\,mas (or $\sim$0.70$\pm$0.13\,AU) was inferred by fitting a ring model with a ring width of 20\% of the inner ring radius.

To get reliable line visibility values, we derived the continuum-corrected (hereafter pure) line visibility only in the spectral regions where the line flux is higher than $\sim$10\% of the continuum flux.
The results are shown in Figure~\ref{fig:line_v}. 
The average line visibility is $\sim$1$\pm$0.2 for the shortest baseline ($\sim$44\,m), $\sim$0.9$\pm$0.15 for the medium baseline ($\sim$61\,m), and $\sim$0.8$\pm$0.1 for the longest baseline ($\sim$89\,m).
For this baseline, the line is spatially resolved even when the error bars are taken into account.

The size of the line-emitting region was obtained by fitting a circular symmetric Gaussian model to the line visibilities presented in Fig.~\ref{fig:line_v}. 
We obtain a Gaussian HWHM radius of 0.65$\pm$0.09\,mas or $\sim$0.29$\pm$0.04\,AU at a distance of 440\,pc.
Similarly, by fitting a ring model with a ring width of 20\% of the inner ring radius, an inner radius of 0.70$\pm$0.09\,mas, or $\sim$0.31$\pm$0.04\,AU, is inferred.

\section{Disc and disc-wind models}
\label{model:sec}

To constrain the physical processes taking place in the innermost disc regions, a proper
physical modelling is required.
As in our previous interferometric works~\citep[see][]{weigelt,rebeca15}, 
to analyse the observed Br$\gamma$ line profile, line visibilities, differential phases, and 
closure phase we employed a previously developed disc-wind (DW) model (see Sect.~\ref{discwind_model:sec}) 
and developed a continuum disc model; both models are adapted to the HD 98922 stellar parameters and observations.

\subsection{Disc-wind model}
\label{discwind_model:sec}

A detailed description of our DW model, its parameters, and the algorithms used for the model computation
can be found in our previous papers~\citep[see][]{grinin11,weigelt,tambovtseva,rebeca15}. 
In the following, we briefly outline the model's main characteristics and define its free parameters, which are listed in Table~\ref{model_parameters:tab}.

The developed model is a warm disc-wind model~\citep[see e.g.][]{safier,garcia01a}, including only hydrogen atoms with constant temperature ($\sim$10\,000\,K).
The wind is rapidly heated by ambipolar diffusion to a temperature of $\sim$10\,000\,K, and  
the wind electron-temperature ($T_{\rm e}$) in the acceleration zone near the disc surface is not high enough to excite the Br$\gamma$ line emission, which is emitted further out along the wind stream lines.

Briefly, the disc wind is launched from a disc inner radius $\omega_1$ to an outer radius $\omega_N$ (called wind footpoints),
where $N$ is the number of streamlines. 
Its half opening angle ($\theta$) is defined as the angle between the innermost wind streamline and the system axis. 
The local mass-loss rate per unit area on the disc 
surface is defined as $\dot{m}_{w}(\omega)\sim \omega^{-\gamma}$, where $\gamma$ is the mass-loading parameter that controls the ejection efficiency. 
Therefore, the total mass-loss rate ($\dot{M}_w$) is
\begin{equation}
\dot M_{w} = 2\int\limits_{\omega_1}^{\omega_N}\dot
m_{w}(\omega)\,2\,\pi\,\omega\,d\omega.
\end{equation}

Finally, the $\beta$ parameter in Table~\ref{model_parameters:tab} regulates the acceleration of the wind along the streamlines~\citep[see][]{tambovtseva}.
Some model parameters such as $\dot{M}_w$, the stellar parameters, and the disc inclination angle were derived in this work or were taken from the literature 
(see Tab.~\ref{stellar_parameters:tab}). 
The  explored range of values of $\dot{M}_w$ varies from $\sim$0.05 to 0.3 times the inferred mass accretion rate ($\dot{M}_{acc}$, see Table\,\ref{stellar_parameters:tab}). 
The value for the corotation radius ($R_{cor}$\footnote{with $R_{corr}$=(GM$_*$/$v_{\omega}^2$)$^{1/3}$, where G is the gravitational constant,
M$_*$  the stellar mass, and $v_{\omega}$  the stellar angular velocity.}) obtained from the $v_{rot}\,sin\,i$ value (Table\,\ref{stellar_parameters:tab}) is $R_{cor}$=1.55\,$R_{*}$.
The estimated value of $R_{cor}$ is then set as a lower limit to the inner disc-wind launching radius value $\omega_1$ to allow for the presence of a hypothetical 
magnetosphere within this region.

To find the best model that reproduces our observations, we adopt the following steps~\citep[for more details, see also][]{weigelt,grinin11}.
First, we fit the observed Br$\gamma$ line profile and intensity. To this end, we solve the equations of statistical equilibrium and compute the population of the \ion{H}{i} atomic 
levels and the ionisation degree along each streamline using the numerical codes developed by \citet{grinin90} and \citet{tambovtseva01} for moving media.
Second, we calculate the intensity of the line radiation in the entire emitting volume filled in by the disc wind and compare both modelled and observed line profiles. 
When a good agreement is found, we compute the brightness distribution of the disc wind and the disc continuum at different radial velocities,
i.e. a two-dimensional intensity distribution map, which provides us with the interferometric observables. 
The modelling results are, in this way, directly compared with the observations until the best-fit model is found.
The best model is that which best matches, within the error bars, all 
eleven interferometric observables (line profile  shape and intensity, visibilities, pure line visibilities, differential phases, and closure phase)
of our observation within the selected range of parameters listed in Table~\ref{model_parameters:tab}. 

This latter reports the free model parameters, the explored range of parameters (Column\,2), 
and the values of our best-fit model, called P5 (Column\,3 of Tab.~\ref{model_parameters:tab}).
P5 model was selected after checking hundreds of different models configurations (see Column\,2, Tab.~\ref{model_parameters:tab}).

\begin{table*}
\centering
 \caption[]{\label{model_parameters:tab} Disc-wind accretion model parameters: range of explored values and values for our
 best-fit model P5}
\begin{tabular}{lcc}
 \hline \hline\\[-5pt]
Parameter\tablefootmark{a} &  Range of Values & Model P5 \\
 \hline\\[-5pt]
Half opening angle ($\theta_1$)     &   20$\degr$--60$\degr$ & 30$\degr$ \\
Inner radius - $\omega_1$(R$_*$)    &   2--3 (0.07--0.1\,AU) & 3 (0.1\,AU)  \\
Outer radius - $\omega_N$(R$_*$)    &   4--30 (0.13--1\,AU) & 30 (1\,AU)  \\
Acceleration parameter ($\beta$)    &   3--7     &   5     \\
Mass load parameter ($\gamma$)      &   2--5      &   3      \\
Mass-loss rate - $\dot{M}_w$(M$_\sun$\,yr$^{-1}$) & 10$^{-8}$--3$\times$10$^{-7}$ & 2$\times$10$^{-7}$ \\
\hline
\end{tabular}
\tablefoot{\tablefoottext{a}{See \citet{weigelt} for definition and detailed description of the listed disc-wind model parameters}
}
\end{table*}


\subsection{Continuum disc model}
\label{disc_model:sec}

To reproduce our AMBER HR interferometric observations, we have assumed
that the HD 98922 continuum emission consists of three different components: the star,
an inner disc, and an outer dusty disc~\citep[see e.g.][]{tannirkulam08a,dullemond}.

\begin{figure}[h]
\centering
\includegraphics[width=7cm]{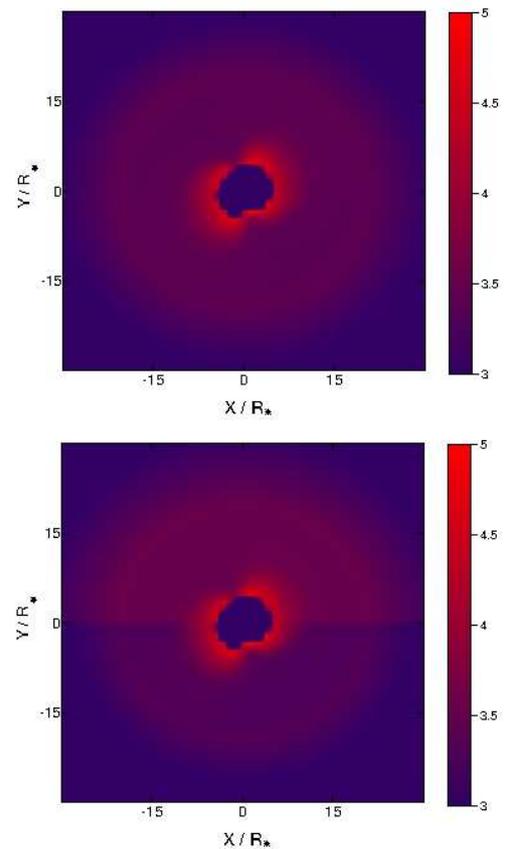}
\caption{\label{models:fig} {\it Upper panel}: Br$\gamma$ intensity distribution map
(for $v_{r}$=0\,km\,s$^{-1}$) of the best-fit disc-wind model (P5) overlapped on the map of the
disc continuum with symmetric brightness distribution. X and Y coordinates are in $R_{*}$. 
{\it Lower panel}: Same as in the upper panel, but for the disc continuum with asymmetric brightness distribution.}
\end{figure}

The main constraints for our developed temperature-gradient model of the
continuum emission are provided by: \textit{i)} the F$_{obs}(K)$/F$_{*}(K)$ 
ratio, namely the ratio between the HD 98922 observed flux continuum in the $K$-band (star plus accretion disc) and
the theoretical stellar flux, and \textit{ii)} the absolute visibilities of the continuum.
The stellar continuum at the Br$\gamma$ rest wavelength (F$_{*}(K)$) was
computed from a Kurucz synthetic spectrum with $T_{\mathrm{eff}}$=10\,400\,K and $log\,g$=3.5 
(see Sect.~\ref{physpars:sec} and Tab.~\ref{stellar_parameters:tab}), whereas the F$_{obs}(K)$
is computed from the YSO $K$-band magnitude (see Tab.~\ref{stellar_parameters:tab}).
The adopted disc inclination angle is the same as for the disc-wind model ($i$=20$\degr$, see Tab.~\ref{stellar_parameters:tab}).
To fit the disc continuum level of our observations, we use an inner gaseous disc plus an outer 
dusty disc, as in \citet{tannirkulam08b} and \citet{benisty10}. The sum of fluxes from the gaseous and dusty disc matches 
the observed flux close to the Br$\gamma$ wavelength. The inner disc has constant brightness distribution and
ranges from $R_{in}$=4\,$R_*$ (or 0.14\,AU) to $R_S$=20\,$R_*$ (or 0.7$\pm$0.2\,AU).
The dust temperature ($T_{\mathrm{d0}}$) at
this radius ($R_S$) is 1500\,K, that is, the silicate dust sublimation temperature~\citep[see e.g.][]{dullemond}. 
The dust temperature exponentially decreases with
the distance $r$ from the star accordingly with the following power law: $T_d(r) = T_{d0}
(r/R_S)^\alpha$,  with $\alpha$=-0.5. 
The intensity of the radiation in the continuum was calculated in the blackbody approximation. 
The disc brightness is distributed symmetrically along azimuthal direction.
Fig.~\ref{models:fig} (upper panel) shows the disc continuum model plus the Br$\gamma$ intensity distribution map
(for $v_{r}$=0\,km\,s$^{-1}$) of the best-fit disc-wind model (P5). 

\subsection{Modelling results}
\label{sect:modelresults}

As seen in Sect.~\ref{AMBER_results:sec}, the Br$\gamma$ emission line shows V-shaped differential phases 
and slightly asymmetric visibilities (see Fig.~\ref{fig:total_v}).

\begin{figure}
\centering
\includegraphics[width=8cm]{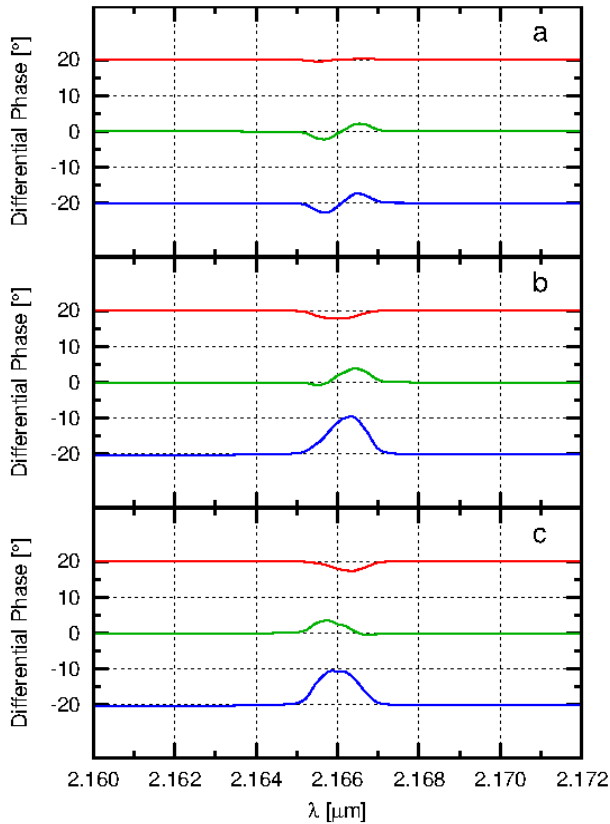}
\caption{\label{dphase:fig} Differential phases for the three observed baselines (as in Fig.~\ref{fig:total_v}), calculated for the best disc-wind model P5 plus: symmetric disc model
(\textit{a}, upper panel), asymmetric disc model with a counter clockwise (\textit{b}, middle panel), and clockwise (\textit{c}, lower panel) disc and disc-wind
rotation. For clarity, the differential phases of the first and last baselines are shifted by +20$\degr$ and -20$\degr$, respectively.}
\end{figure}

Because of the aforementioned shape of the differential phases, our first attempts at modelling these interferometric observables with our disc-wind model plus the continuum disc model described 
in Sect.~\ref{disc_model:sec} were unsuccessful (see Fig.~\ref{symmdisc:fig} in Appendix~\ref{appendix-b:sec}). Indeed, the combination of these disc and disc-wind models produces, 
as expected, symmetric S-shaped differential phases (originating from disc rotation, see Figure~\ref{dphase:fig}, upper panel) and symmetric visibilities in the Br$\gamma$ line.

In principle, the unusual shapes of the observed visibilities and differential phases, which are a measurement of the photocentre shifts, 
can be caused by an asymmetric Br$\gamma$ line-emitting region, an asymmetric continuum-emitting region, or by
a combination of both. As a consequence, the differential phase in the blue or red wings of the Br$\gamma$ can be displaced, and the visibilities become asymmetric.
An asymmetric Br$\gamma$ line-emitting region could be a plausible explanation for the observed differential phase. In theory, this might originate from a collimated wind/jet,
whose red-shifted lobe is hidden by the disc~\citep[see e.g.][]{benisty10}. However, in this case, we would observe a blue-shifted photocentre shift, whereas our observations clearly
show a red-shifted peak in the differential phases. On the other hand,
it is worth noting that the HD\,98922 NIR continuum emission shows an asymmetric brightness distribution on small and large
spatial scales (from $\sim$1\,AU to tens of AUs), as observed in both VLTI/PIONIER and SINFONI data~\citep[see Sect.~\ref{SINFONI-results:sec}, and][]{kluska}.
Therefore, an asymmetric continuum might be responsible for the observed shape in the differential phases and visibilities.

To verify this hypothesis, we modified the disc model of Sect.~\ref{disc_model:sec}, developing a very simple asymmetric continuum disc model, 
which mimics the difference in brightness between the two halves of the disc. 
However, it is beyond the scope of this paper to model the observed disc asymmetries in details and at small scales. 
Therefore, for the sake of simplicity, the intensity of the radiation was increased by a factor $f_c$ in the northern half
of the disc and decreased by the same factor in the southern half of the disc of both gaseous and dusty disc regions.
As a consequence, the disc brightness from PA 270$\degr$ to 90$\degr$ (i.e. the northern half of the disc) 
is brighter than the southern half of the disc (extending from PA 90$\degr$ to 270$\degr$).
Lower panel of Figure.~\ref{models:fig} shows the asymmetric disc continuum model plus the Br$\gamma$ intensity distribution map
(for $v_{r}$=0\,km\,s$^{-1}$) of the best-fit disc-wind model (P5).
The brightness contrast ($f_c$) between the two halves of the disc is a free parameter of the model. 
The asymmetric disc was then rotated by 10$\degr$ steps until we obtained
a solution that best fits both visibilities and differential phases at the three observed
baselines. In order to fit the continuum level of the three
baselines, we changed the $f_c$ value, which was varied from 1 to 4. 
The best continuum model has an $f_c$ value of 2.5 and a disc rotation of 70$\degr$.
Although very simple, our asymmetric disc model, combined with the disc-wind model described in Sect.~\ref{discwind_model:sec}, 
is able to reproduce the V-shaped differential phases (see Fig.~\ref{dphase:fig}, middle panel).
Middle and lower panels of Fig.~\ref{dphase:fig} present differential phases for different rotation of
the disc and the disc wind: counter clockwise (\textit{b}; middle panel) and clockwise rotation (\textit{c}; lower panel). It should
be noted that the shapes of the differential phases in the two cases are slightly different.
The reason is the disc wind. By modifying the direction of the disc-wind rotation, the
distribution of the disc-wind brightness changes but the distribution of the disc brightness
remains unchanged. Thus, the differential phases are sensitive to the direction of the disc
and disc-wind rotation.

\begin{figure}
\centering
\includegraphics[width=8cm]{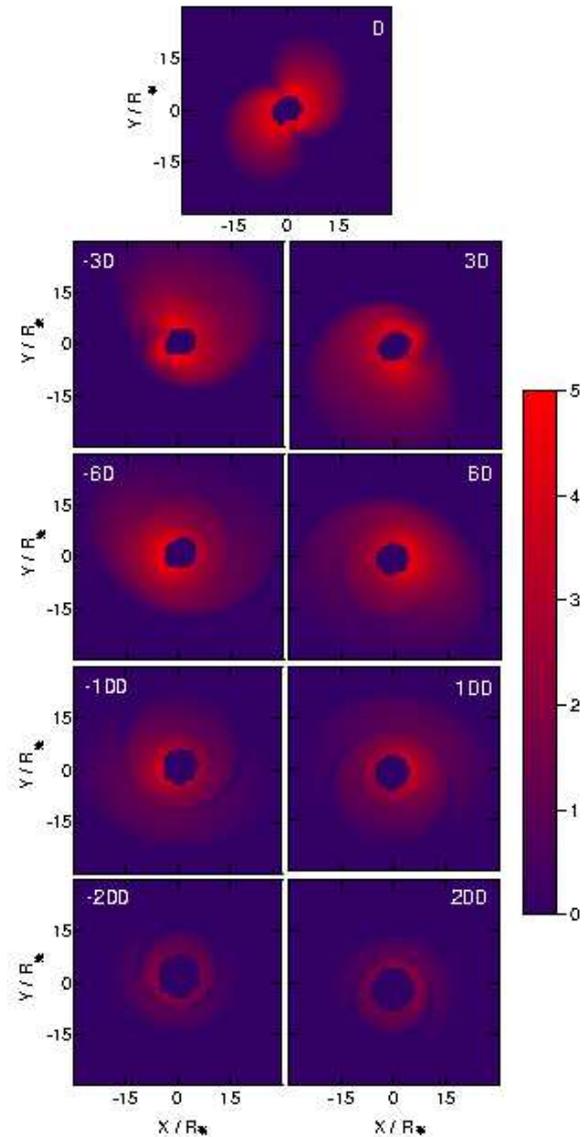}
\caption{\label{disc_wind_model:fig} Br$\gamma$ intensity distribution maps of our best disc-wind model
P5 (see values of parameters in Table\,\ref{model_parameters:tab}, Column\,3) in logarithmic scale (arbitrary units).
Each panel shows the intensity map for a different radial velocity, indicated by a white label
in units of km\,s$^{-1}$. X and Y coordinates on each map are in $R_{*}$.
The system axis inclination angle with respect to the line of sight is assumed to be 20$\degr$ (see Table\,\ref{stellar_parameters:tab}). 
Continuum emission from the disc and central star, located at coordinates 0,0, are not shown.}
\end{figure}

\begin{figure}[h!]
        \centering
{\includegraphics[width= 14cm, angle =0]{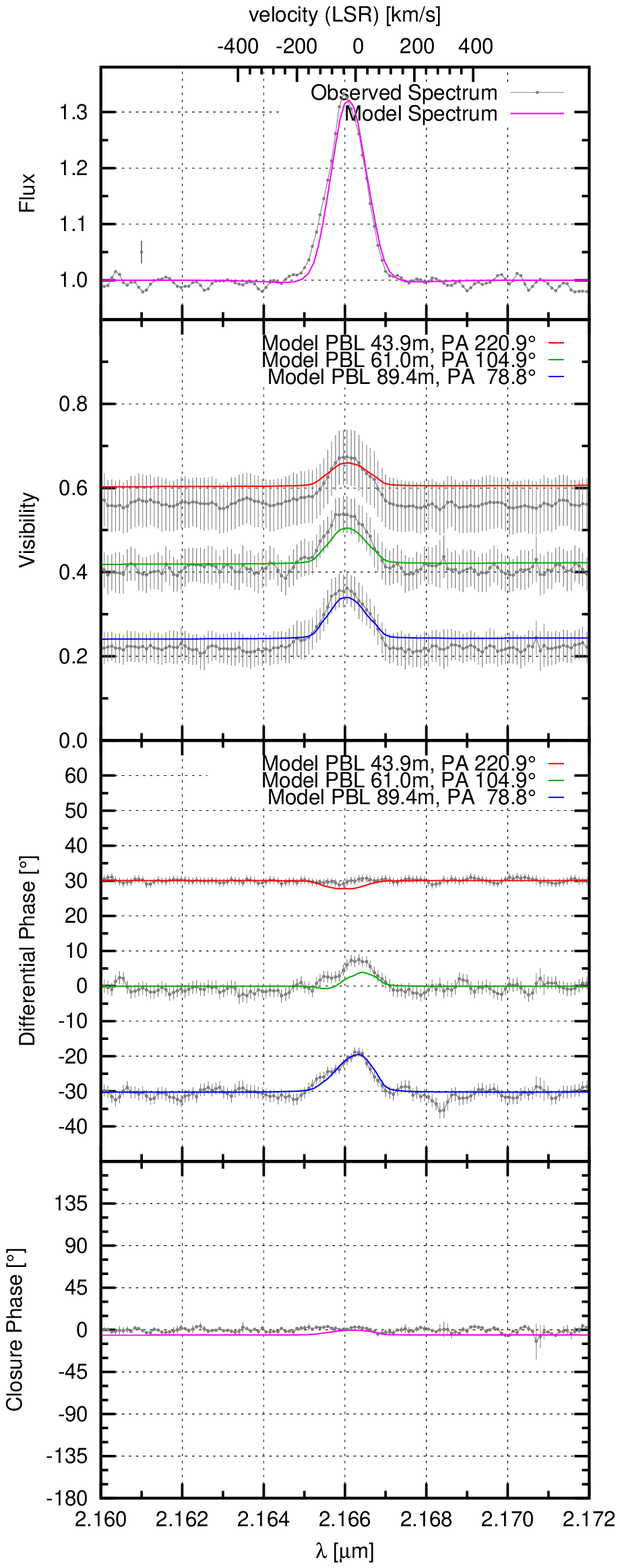}}
        \caption{\label{fig:model} Comparison of our interferometric observations with the interferometric
observables derived from our best disc-wind model P5 plus asymmetric disc model.
From top to bottom, observed Br$\gamma$ line profile (grey) and model line
profile (pink), observed visibilities (grey dots with error bars) and model
visibilities (coloured lines), observed and model wavelength-differential phases, and observed and model closured phase.}
\end{figure}

Figure~\ref{disc_wind_model:fig} shows the Br$\gamma$ intensity distribution maps of our best disc-wind model (P5) in logarithmic scale, 
as a function of different Br$\gamma$ radial velocities. 
The adopted system axis inclination angle with respect to the line of sight is 20$\degr$~\citep[see Table~\ref{stellar_parameters:tab};][]{hales}.
Column\,3 of Table\,\ref{model_parameters:tab} reports the values of the model free parameters.
Figure~\ref{fig:model} shows a comparison of our best model with the interferometric observables.
Despite the simplicity of the proposed disc model, the interferometric observables are approximatively well reproduced, demonstrating that 
V-shaped differential phases and slightly asymmetric line visibilities may originate from the uneven brightness distribution of the disc.
It is also worth to note that the value derived for $f_c$ (i.e. 2.5) is, within the error bars, identical to the average brightness ratio between the northeastern and southwestern disc 
regions of the SINFONI spectral images (see Sect.~\ref{SINFONI-results:sec}), and that the PA of our asymmetric disc model (i.e. 70$\degr$) is compatible with
the observed asymmetric disc orientation in Fig.~\ref{fig:disc_asymmetry}.
In Fig.~\ref{fig:line_v}, we compare observational and model results of the Br$\gamma$ pure line visibilities (pink and black solid lines, respectively).
Our model approximatively reproduces the observed values at the shortest and medium baselines (upper and central panels), but it slightly overestimates the pure line visibility
at the longest baseline (lower panel). This result represents a good compromise.
To obtain a lower value of the modelled pure line
visibility at the longest baseline, it would be necessary to increase
the size of the disc-wind region (i.e., the value $\omega_N$). On the other hand this would
increase the values of the differential phase, and the resulting model would not match our observations.

Moreover, we also note that the Br$\gamma$ emitting region is spatially extended. As a consequence, it cannot be modelled with a compact, spatially unresolved component alone, 
such as a magnetosphere~\citep[size $<$2-3\,R$_*$; see e.g.][]{tambovtseva,rebeca15}, that cannot be spatially resolved at these baselines. Therefore we conclude that the Br$\gamma$
emission must be at least a combination of a compact (i.e. spatially unresolved) emission (magnetosphere) plus an extended spatially resolved  
emission (i.e. the disc wind). The addition of a compact region (i.e. visibility value = 1 at all baselines) implies that the size of the spatially resolved component must be enlarged to match
the observed visibility values. This can be achieved with a larger disc-wind footpoint radius $\omega_N$, which, in turn, gives larger values
of the differential phases. In addition, changing the footpoint radius modifies the line profile as well.
Thus our model has to take into account these three observables (visibility, differential phase, and line profile).
In conclusion, by comparing our model with the observations, we do not rule out that some of the Br$\gamma$ emission may originate from the magnetosphere (i.e. a spatially unresolved region), 
however we can exclude that this latter is the main mechanism, which produces the Br$\gamma$ emission~\citep[see also][]{rebeca15}.

Finally, it is worth  comparing the mass-loss rate derived from our model with the mass accretion rate inferred from the observations, as their ratio ($\dot{M}_w$/$\dot{M}_{acc}$)
usually provides constraints on the efficiency of the accretion/ejection mechanism in young stars, typically ranging from 0.1 to 0.3~\citep[see e.g.][]{shu98,ferreira06}.
In our case, the accuracy of the $\dot{M}_{acc}$ value is $\sim$30\%. This uncertainty mainly originates from the employed empirical relationship~\citep[see e.g.][]{mendigutia11,mendigutia13}.
On the other hand, it is more difficult to estimate the accuracy of the $\dot{M}_w$ value as inferred from our best-fit model, but it is probably of the same order as the  $\dot{M}_{acc}$ value,
owing to the use of our simplified model of the disc wind and to the uncertainties in the fit of the Br$\gamma$ line profile and intensity.
As a result, their ratio is well within the expected range of values.
It should also be noted that in our best-fit model of HD\,98922 the disc wind starts in the close vicinity of the star.
This can mean that the wind at least partially originates as a result of interaction of the accretion disc with the 
magnetosphere. In this case the $\dot{M}_w$/$\dot{M}_{acc}$ ratio can be higher than that used in the theory of the magneto-centrifugal disc wind (0.1), for example due to the effect of the mechanism 
of the magnetic propeller~\citep[see e.g.][]{illarionov}.

\section{Discussion}
\label{discussion:sec}

\subsection{An asymmetric disc emission} 
\label{asymmetry:sec}

Asymmetries in the disc brightness distribution are not uncommon as they have been observed at NIR wavelengths in several Herbig Ae/Be stars and CTTSs
with the Hubble Space Telescope (HST) or with ground-based telescopes through AO assisted imaging or interferometric imaging~\citep[see e.g.][]{krist,dullemond}.
These images of spatially resolved discs display a wide range of spectacular asymmetries, including arcs~\citep[e.g.][]{fukagawa,krist12}, gaps and dips~\citep[e.g.][]{krist},
warps~\citep[e.g.][]{golimowski,marino}, or more complex morphologies~\citep[e.g.][]{hines,kalas,mazoyer}.

The origin of such asymmetries in discs varies from case to case, but mostly involves
disc geometry, illumination~\citep[e.g.  only one side of the illuminated inner rim is observed; see][]{monnier06}, dust grain composition and distribution, or tidal effects:
\textit{a)} brightness asymmetries between the two halves of the disc may originate from different emission in the forwards/backwards scattered light in inclined discs~\citep[see e.g.][]{krist,dullemond};
\textit{b)} gaps and dips in discs can be produced by shadowing effects caused by optically thick clouds of dust and gas~\citep[see e.g.][]{krist};
\textit{c)} arcs of different brightness and size may originate from shadowing effects, produced by a different inclination between the inner and the outer disc~\citep[see e.g.][]{marino}; 
\textit{d)} arcs can be also produced by dust density variations and stellocentric grain-size segregation;
\textit{e)} arcs and rings can also originate from warped discs produced by tidal interactions~\citep[see e.g.][]{tuthill,demidova,marino}.

To investigate the origin of the asymmetries in the HD 98922 disc intensity distribution, we should analyse
large- and small-scale asymmetries.

Asymmetries on small scales include \textit{i)} asymmetric inner brightness distribution, due to shadowing caused by the vertically
extended structure of the inner rims~\citep[which can appear unusually asymmetric if the disc geometry is not exactly face-on; see e.g.][]{monnier06}, or 
\textit{ii)} brightness peaks and gaps along the disc azimuthal angle.

The large scale asymmetry is represented by the different brightness distribution between the 
NNE and SSW side of the disc (whose ratio is about 2.5). This feature is detected at both VLTI (AMBER and PIONIER)
and SINFONI spatial scales. 

The different emission between forward and backward scattering could be the  simplest explanation.
The fact that the brightness asymmetry lies approximately  in the direction of the system axis also supports this hypothesis.
On the other hand, this asymmetry is observed both at large (SINFONI) and small (VLTI) scales,
and the SINFONI data are likely dominated by scattered light, whereas the VLTI data are possibly dominated 
by thermal emission. These considerations are in favour of an intrinsic perturbation of the disc.

In principle, disc warping is the most plausible explanation.
Recently \citet{demidova} modelled the disc asymmetry of LkHa 101~\citep[a Herbig Ae star; see e.g.][]{tuthill},
which shows a warped disc. Such asymmetry is very similar to that observed in HD98922~\citep[see e.g. Figure~1 in][]{tuthill}.
The warp is generated by the tidal effects of a low-mass companion (with mass ratio $q$ ranging from 0.1 to 0.01) with its orbit slightly inclined with respect to the disc plane. 
The warp height varies as the companion extracts matter out of the  mid-plane of the disc.
These perturbations generate an azimuthal asymmetry of the extinction.
As a result, the illumination of the disc by the star becomes asymmetric.

In the case of HD 98922, the SINFONI data exclude the presence of a companion as close as $\sim$20\,AU from the central source down to $\sim$0.5\,M$_\sun$ 
(see Sect.~\ref{SINFONI-results:sec}, i.e. q$\geq$0.1). Moreover,
the VLTI/PIONIER reconstructed image~\citep[see][]{kluska} does not show any evidence of a close binary (up to a few AUs from the star),
but for these data no sensitivity limit is provided. Indeed, given the VLTI/PIONIER sensitivities~\citep[see e.g.][]{absil}, the dynamical range of the reconstructed image should be lower than 
or equal to 100 (Kluska priv. comm). This roughly gives a lower limit of $H\sim 10.2$\,mag or $M_H \sim 2$\,mag at the adopted distance of 440\,pc.
By adopting the \citet{siess00} evolutionary tracks (see Sect.~\ref{SINFONI-results:sec}), we obtain an upper limit for the mass of any undetected companion
of $\sim$0.4--0.5\,M$_\sun$. Our reasoning implies that the presence of a low-mass object ($M_*\leq 0.5$\,M$_\sun$) or a massive protoplanet cannot be ruled out 
by both SINFONI or VLTI observations. As a consequence the hypothesis of a warped disc, originating from a very low-mass companion (i.e. with mass ratio in the range of 0.1--0.01), still holds.

\subsection{Inner circumstellar structure of HD 98922}

Our interferometric observations show that the Br$\gamma$ line visibilities are higher than those of the continuum, indicating 
that the Br$\gamma$ emitting region is less extended ($\sim$0.3\,AU in radius) than the continuum emitting region ($\sim$0.7\,AU in radius). 
The pure Br$\gamma$ line visibility of HD 98922 is 0.8 at the longest baseline (89\,m),
indicating that the Br$\gamma$ originates from a spatially extended region. 
Our interferometric observables (Br$\gamma$ line profile, visibilities, differential phases, closure phases) are sufficiently well reproduced by a disc-wind model
with a half opening angle of 30$\degr$, wind footpoints extending from $\sim$0.1\,AU to $\sim$1\,AU, and $\dot{M}_w \sim 2 \times$10$^{-7}$\,M$_\sun$\,yr$^{-1}$.

The wind mainly originates from the inner disc region, beyond the corotation radius ($R_{cor}$), which can be placed at $\sim$0.06\,AU from the star
(assuming $\mathrm{v}_{rot}\,sin\,i$=50$\pm$3\,km\,s$^{-1}$ and $i$=20$\degr$, see Table\,\ref{stellar_parameters:tab}).
Notably, the wind model extends slightly beyond the dust sublimation radius, located at $\sim$0.7\,AU.
This is required to fit both the low visibilities and the large differential phases. Indeed, a more compact disc-wind model would produce
small or no differential phases and it would not reproduce our interferometric data~\citep[see][]{rebeca15}.

It is worth  mentioning that some of the previous HR AMBER/VLTI studies of Herbig AeBe stars~\citep[e.g. MWC\,297 and HD\,163296: see][respectively]{weigelt,rebeca15} show similar results to those described here,  
namely that the Br$\gamma$ line is emitted from an extended region that can be modelled with a disc wind.
However, MWC\,297 (SpT=B1V) and HD\,163296 (SpT=A1V) show disc-wind geometries that differ from the one modelled in HD 98922 (SpT=B9V). 
In the first case, the disc wind was entirely launched from a region located in the continuum disc, i.e. well beyond the inner radius of the continuum-emitting disc~\citep[][]{weigelt}.
The disc-wind model of MWC 297 extends from $\sim$17 to 35\,R$_*$ with a large half opening angle of 80$\degr$. 
Moreover, the values of the $\beta$ and $\gamma$ parameters, which regulate the mass load along the streamlines, are quite small (1 and 2, respectively).
This indicates that the mass load is shifted towards the outer streamlines of the wind.
\citet[][]{weigelt} suggest that such features might be caused by the
intense radiation pressure of the central source.
On the other hand, HD\,163296 has a very compact disc wind (extending $\sim$0.16\,AU), confined to the inner gaseous disc~\citep[][]{rebeca15}.
In this case, $\omega_1$ and $\omega_N$ radii are 2 and 4\,R$_*$ with a half opening angle of 45$\degr$. In contrast to MWC\,297 and HD\,98922, HD\,163296 drives a well-collimated
jet~\citep[see e.g.][]{wassell}; therefore, the smaller $\omega$ values might be related to this.
The modelled disc wind of HD\,98922 has an intermediate geometry, with the Br$\gamma$ line-emitting region located mostly in the inner disc, 
but extending slightly beyond the dust sublimation radius.
Our disc-wind  model extends from 3 to 30\,R$_*$ with a half opening angle of 30$\degr$. Although spectroscopy indicates a strong outflow originates from its disc~\citep[e.g. this work; see also][]{grady},
no signature of any collimated jet has been detected so far. On the other hand, the $\beta$ and $\gamma$ parameters are identical to those of HD\,163296, suggesting that the mass
load is similarly distributed along the stream lines, but along a larger portion of the disc.

Following these considerations, our modelling seems to suggest that different types of outflows/jets can be modelled with different disc-wind geometries. 
Moreover, these disc-wind geometries seem to display an evolutionary trend, following the  different spectral types and/or evolutionary stages of the central sources.
It is also tempting to speculate on some additional correlations between, for example, the Br$\gamma$ size, its location, and the stellar type and mass of the central source. 
If our previous speculations are correct, we would then expect that disc wind geometry, kinematics, and Br$\gamma$ size depend on the central source, namely its mass and/or evolutionary stage.
Nevertheless, it is also clear that the sample so far studied is still too limited to draw any firm conclusion.

\section{Conclusions} 
\label{conclusion:sec}

In this paper, we analyse the main physical parameters and the circumstellar environment of the young Herbig Be star
HD 98922. We present AMBER high spectral resolution ($ \mathrm R $=12\,000) interferometric observations across the Br$\gamma$ line, as well
as UVES (high-resolution UVB spectroscopy) and SINFONI-AO assisted (NIR integral field spectroscopy) ancillary data.
To interpret our observations, we also developed a magneto-centrifugally driven disc-wind model along with a very simple asymmetric continuum disc model.  
The main results of this work are the following:

\begin{enumerate}
\item[-] UVES high-resolution spectrum displays \ion{H}{i} photospheric absorption lines from the Balmer series (from \ion{H}{11} to H$\beta$) with superimposed 
emission originating from circumstellar activity. Additional weaker photospheric absorption features from atomic lines (\ion{He}{I}, \ion{Ca}{II}, \ion{Fe}{I}, \ion{Fe}{II},
\ion{Mg}{I}, \ion{Mg}{II}, \ion{Si}{I}, \ion{Si}{II}) and circumstellar emission lines (\ion{Fe}{II} and \ion{Ti}{II}) are also detected in the spectrum.

\item[-] Our analysis of the UVES spectrum indicates that HD 98922 is a young (5$\times$10$^5$\,yr) Herbig Be star (SpT=B9V), located at a distance of 440$\pm^{60}_{50}$\,pc, 
with $\dot{M}_{acc}$=(9$\pm$3)$\times$10$^{-7}$\,M$_{\sun}$\,yr$^{-1}$.

\item[-] SINFONI $K$-band AO-assisted imaging shows a spatially resolved circumstellar disc ($\sim$140\,AU in diameter) with asymmetric brightness distribution. 
In particular the NNE side is $\sim$2.5 times brighter than the SSW side.
Other features at smaller spatial scales include peaks and dips along the disc. 

\item[-] VLTI/AMBER high spectral resolution observations ($ \mathrm R $=12\,000) allow us to study the interferometric observables
(i.e.  line profile, visibilities, differential phases, and closure phases) in many spectral channels across the Br$\gamma$ line.
Differential phases are V-shaped and line visibilities are slightly asymmetric.

\item[-] Our interferometric observations indicate that the Br$\gamma$ line visibilities are higher than those of the continuum, i.e.
the Br$\gamma$ emitting region is less extended than the continuum emitting region. 
For the first time, HD 98922 continuum-corrected Br$\gamma$ pure line visibilities are 0.8 at the longest baseline (89\,m),
indicating that the Br$\gamma$ originates from a spatially extended region. By fitting geometric Gaussian and ring models to the derived
pure line visibilities, we infer a HWHM Gaussian radius of 0.65$\pm$0.09\,mas (or $\sim$0.29$\pm$0.04\,AU at a distance of 440\,pc) and ring-fit radius of 0.70$\pm$0.09\,mas ($\sim$0.31$\pm$0.04\,AU)
for the Br$\gamma$ emitting region.
The Br$\gamma$ line radius is smaller than the inner rim radius of 0.7\,AU of our continuum temperature model. 

\item[-] To obtain a more physical interpretation of our Br$\gamma$ AMBER observations, we employed our own line radiative transfer disc-wind model 
\citep[see][]{weigelt,grinin11,tambovtseva,rebeca15}. We computed a model that approximately reproduces all the interferometric observables. As a result,
our modelling suggests that the observed Br$\gamma$ line-emitting region mainly originates from a disc wind with a half opening angle of
30$\degr$, wind footpoints  extending from $\sim$0.1\,AU to $\sim$1\,AU, and $\dot{M}_w$=2$\times$10$^{-7}$\,M$_\sun$\,yr$^{-1}$.

\item[-] The observed V-shaped differential phases and slightly asymmetric visibilities can be reproduced by combining a simple asymmetric continuum disc model
with our Br$\gamma$ disc-wind model.

\end{enumerate}

\begin{acknowledgements}
We thank an anonymous referee for his/her comments, which improved the paper.
We are also grateful to Jaques Kluska for providing us with information about the VLTI/PIONIER observations
reported in Kluska et al. 2014.
A.C.G. and R.G.L. were partially supported by the Science Foundation of Ireland, grant 13/ERC/I2907.
L.V.T. was partially supported by the Russian Foundation for Basic Research (Project 15-02-05399).
S.K. acknowledges support from an STFC Ernest Rutherford fellowship (ST/J004030/1), Ernest Rutherford
Grant (ST/K003445/1), and Marie Curie CIG grant (SH-06192). 
V.P.G. was supported by grant of the Presidium of RAS P41.
This research has also made use of NASA's Astrophysics Data System Bibliographic Services and the SIMBAD database, operated
at the CDS, Strasbourg, France.
\end{acknowledgements}

\bibliographystyle{aa}
\bibliography{references}

\Online
\begin{appendix}
\section{AMBER interferometric results  26  December 2012}
\label{appendix:sec}

\begin{figure}
        \includegraphics[width= 14cm, angle =0]{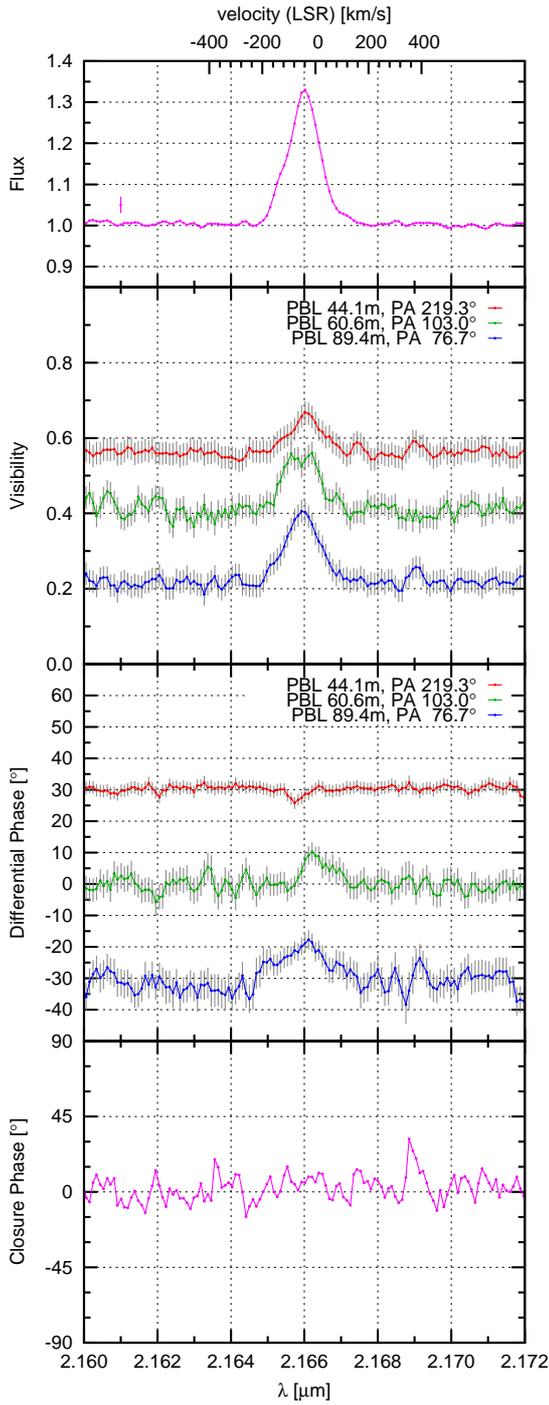}
                \caption{\label{fig:total_v2} AMBER observation of HD~98922 at spectral resolution of $ \mathrm R $=12\,000 (26 Dec 2012). 
                From top to bottom: wavelength dependent flux, visibilities, wavelength-differential phases, and closure phases observed at three different projected baselines (see labels in figure).
                For clarity, the differential phases of the first and last baselines are shifted by +30$\degr$ and -30$\degr$, respectively.
                }

\end{figure}

\newpage
\section{Modelling results for disc-wind model P5 and symmetric disc model}
\label{appendix-b:sec}
\begin{figure}[h!]
        \centering
{\includegraphics[width= 14cm, angle =0]{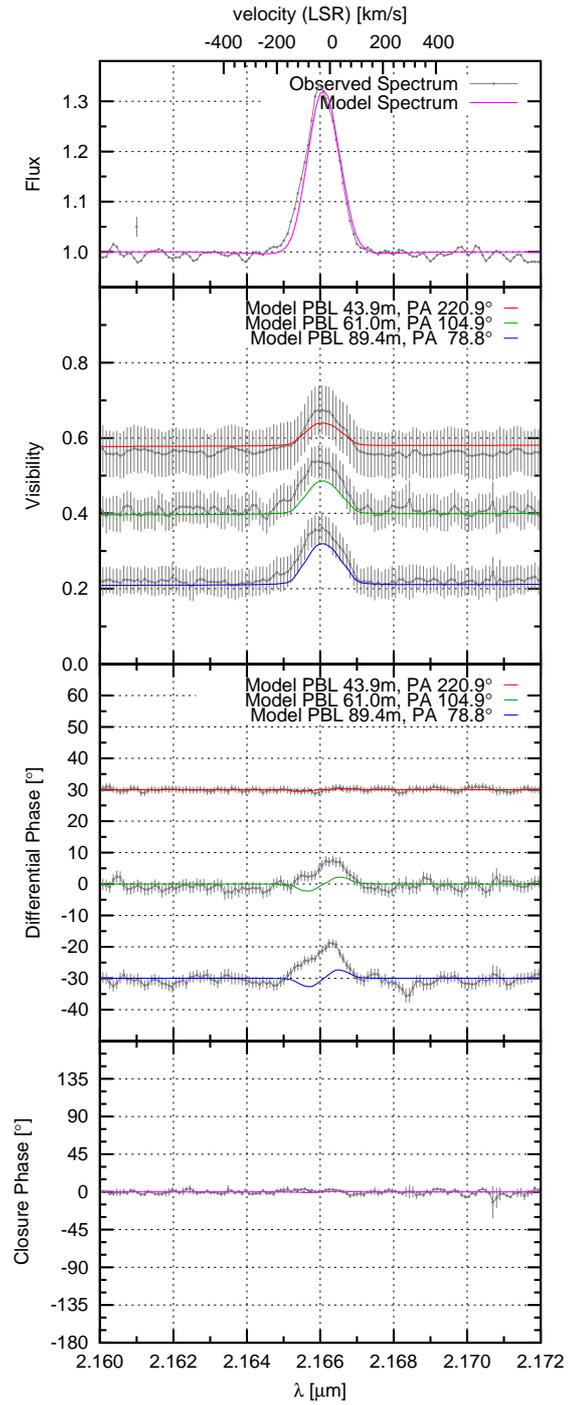}}
        \caption{\label{symmdisc:fig} Comparison of our interferometric observations with the interferometric
observables derived from our best disc-wind model P5 plus the symmetric disc model described in Sect.~\ref{disc_model:sec}.
From top to bottom, observed Br$\gamma$ line profile (grey) and model line
profile (pink), observed visibilities (grey dots with error bars) and model
visibilities (coloured lines), observed and model wavelength-differential phases, and observed and model closure phase.}
\end{figure}

\end{appendix}
\end{document}